\documentclass[preprint,journal]{vgtc}       

\ifpdf
  \pdfoutput=1\relax                   
  \usepackage{graphicx}                
  \DeclareGraphicsExtensions{.pdf,.png,.jpg,.jpeg} 
\else
  \usepackage{graphicx}                
  \DeclareGraphicsExtensions{.eps}     
\fi%

\graphicspath{{figures/}{pictures/}{images/}{./}} 

\usepackage{microtype}                 
\PassOptionsToPackage{warn}{textcomp}  
\usepackage{textcomp}                  
\usepackage{mathptmx}                  
\usepackage{times}                     
\usepackage{cite}                      
\usepackage{tabu}                      
\usepackage{booktabs}                  
\usepackage{color}

\usepackage{amsmath}
\usepackage{amssymb}
\usepackage{amsfonts}
\usepackage{subfig}
\usepackage{floatrow}
\usepackage{bm}
\usepackage{helvet}
\usepackage[font={scriptsize,sf},tableposition=top]{caption}
\usepackage{float}
\floatstyle{plaintop}
\restylefloat{table}
\usepackage[keeplastbox]{flushend}

\usepackage[locale=FR]{siunitx} 
\usepackage{tikz}
\usepackage{pgfplots}
\usepackage{multicol}
\preprinttext{}

\onlineid{1203}

\vgtccategory{Research}

\vgtcpapertype{algorithm/technique}

\title{
An Incremental Dimensionality Reduction Method for Visualizing Streaming Multidimensional Data
}

\author{Takanori Fujiwara, Jia-Kai Chou, Shilpika, Panpan Xu, Liu Ren, and Kwan-Liu Ma}
\authorfooter{
\item
 Takanori Fujiwara, Jia-Kai Chou, Shilpika, and Kwan-Liu Ma are with University of California, Davis. E-mail: \{tfujiwara, jkchou, fshilpika,klma\}@ucdavis.edu.
\item
 Panpan Xu and Liu Ren are with Bosch Research North America. E-mail: \{panpan.xu, liu.ren\}@us.bosch.com.
}

\shortauthortitle{Fujiwara \MakeLowercase{\textit{et al.}}: An Incremental Dimensionality Reduction Method for Visualizing Streaming Multidimensional Data}

\abstract{
Dimensionality reduction (DR) methods are commonly used for analyzing and visualizing multidimensional data. However, when data is a live streaming feed, conventional DR methods cannot be directly used because of their computational complexity and inability to preserve the projected data positions at previous time points. In addition, the problem becomes even more challenging when the dynamic data records have a varying number of dimensions as often found in real-world applications. This paper presents an {\em incremental} DR solution. We enhance an existing incremental PCA method in several ways to ensure its usability for visualizing streaming multidimensional data. First, we use geometric transformation and animation methods to help preserve a viewer's mental map when visualizing the incremental results. Second, to handle data dimension variants, we use an optimization method to estimate the projected data positions, and also convey the resulting uncertainty in the visualization. We demonstrate the effectiveness of our design with two case studies using real-world datasets.
}


\CCScatlist{
 \CCScat{I.3.8}{Computer Graphics}{Applications}
}

\vgtcinsertpkg

\definecolor{red}{rgb}{0,0,0}

\ieeedoi{10.1109/TVCG.2019.2934433}

\begin{document}

\firstsection{Introduction}
\maketitle

\setlength{\abovedisplayskip}{2pt}
\setlength{\belowdisplayskip}{2pt}

In support of effective analysis, studying how to best represent data with high dimensionality has been one of the major focuses in the visualization community~\cite{liu2014survey}. 
Techniques such as parallel coordinates~\cite{inselberg1987parallel}, scatterplot matrices~\cite{hartigan1975printer}, and dimensionality reduction (DR)~\cite{van2009dimensionality}, were developed to present multidimensional structural information of data as a projection onto lower-dimensional space~\cite{liu2017visualizing} (typically 2D). 
Compared to the other two methods, DR's visual results allow us to discern relationships between data records more easily.
Also, DR is less prone to visualization scalability issues in terms of both visual representation and data dimensionality. 
Because of these advantages, DR is commonly used for an initial exploration of multidimensional data in many fields, such as biology~\cite{hollt2018cyteguide}, social science~\cite{tsai2011dimensionality}, and machine learning~\cite{rauber2017visualizing}.

In many streaming data applications, including social media text-analysis~\cite{bosch2013scatterblogs2}, traffic flow monitoring~\cite{cao2018voila}, financial fraud detection~\cite{webga2015discovery}, computer network screening~\cite{arendt2016cyberpetri,xia2013online}, and assembly lines performance diagnostics~\cite{xu2017vidx}, the data is often multidimensional. 
For these datasets, utilizing effective visualizations is crucial for performing timely analysis.
However, applying DR to continuously updating dataset is not a trivial task due to the following challenges: (1) the computation time needs to keep up with the data-collection rate, (2) the viewer's mental map needs to be preserved, and (3) the potentially non-uniform number of dimensions for each data point needs to be handled. 

The computational cost is the primary concern when using DR for streaming data.
As new data keeps coming in, the time for calculating and updating positions of data points must be fast enough to keep the visualization up-to-date. 
This becomes particularly difficult when the number of data records and/or the number of dimensions is large. 

Another challenge is how to preserve a viewer's mental map while continuously updating the visualization from DR results. 
Most of the well-known DR methods, such as principal component analysis (PCA)~\cite{jolliffe1986principal}, multidimensional scaling (MDS)~\cite{torgerson1952}, and t-distributed stochastic neighbor embedding (t-SNE)~\cite{maaten2008visualizing}, originally designed their approach for a static setting. 
As a result, each time DR is directly applied to a streaming data, the projected data points' positions could look drastically different from the positions obtained at the previous time step. 
This would, therefore, easily interrupt the viewer's analysis process and be too difficult to maintain a mental map.

The last, and perhaps, the most challenging problem in employing DR for streaming data analysis is caused by the non-uniform number of dimensions. 
In scenarios where an analysis is based on multiple data sources, some of the data points could have missing features if those features have not been recorded yet.
For example, when monitoring a product assembly line, we may measure the time it takes for products to pass through the work stations that assemble the products, and then use the measured time as features for further analysis.
However, at any given time point, some products might have already been assembled (i.e., the full set of features is collected), while the others are still being processed at one of the work stations (i.e., at least one of the features is missing). 
In such cases, ordinary DR methods are not directly applicable as they cannot handle data records with a variant number of dimensions. 

To address the challenges mentioned above, we introduce a method for visualizing DR results for streaming data. 
While there are many different types of DR methods, in this paper, we focus on PCA because of its popularity for visualization~\cite{sacha2017visual}.
To reduce the amount of computation needed at each iteration, we employ incremental PCA~\cite{ross2008incremental}, which calculates the new results by using the results obtained from the previous step as a base and then updates according to the newly added information. 
However, the traditional incremental PCA will still rearrange data points' positions at each successive time point. 
We would still then run into the problem of disturbing the viewer's mental map.
In our method, we, therefore, apply a geometric transformation, specifically the Procrustes transformation~\cite{borg1997modern}, to make the transition of each data point's position easier to follow.
In addition, we animate the transition of data points between subsequent time points to reduce the viewer's cognitive load.

To handle data records with a non-uniform number of dimensions, we introduce a position estimation method. 
It estimates where the positions of data points with an incomplete number of features would be in the PCA result of the other data points which would have the full set of features.
We also provide a mechanism to measure the uncertainty introduced by our estimation method. 
By visually presenting uncertainty information, viewers can assess the trustworthiness of the displayed result. 
This will help them make better decisions as well as adjust their hypotheses during the exploration and observation stages of the visualization. 

To present the efficiency of our methods, we conduct performance testing.
The result shows that the calculation time of the methods meets the requirement of supporting real-time applications.
Furthermore, we develop a prototype system integrating our methods to demonstrate their effectiveness with in-depth analysis of real-world datasets. 
Two case studies showcase how our method can be used for visually detecting potential anomalies and finding forming clusters from streaming data.

\section{Related Work}
We survey the relevant works in streaming data visualization and dimensionality reduction methods.

\subsection{Streaming Data Visualization}
\label{sec:streamingVis}
Visualizing streaming data for effective analytics is an important research topic.
Dasgupta et al.~\cite{dasgupta2017human} provided a comprehensive survey of streaming data visualization and its challenges. 
One main challenge is that the visualization needs to be constantly updated with incoming data.
This introduces two major concerns: (1) cognitive load and (2) computational cost. 

Krstajic and Keim~\cite{krstajic2013visualization} summarized the problems related to the cognitive load. 
They compared the occurring changes from streaming data in well-known visualizations, such as scatterplots and streamgraphs, and summarized the potential loss of context from its effects.  
For instance, if a new data value is outside of the current axis range(s) of a scatterplot, we would need to decide whether to update the axis range(s) or not. 
In the case that we decide to make an update, the viewer's mental map then may be lost at the same time. 
On the contrary, if no update is applied, we run into the issue of information loss. 

As for overcoming the issue of computational cost, incremental methods, such as~\cite{tanahashi2015efficient,crnovrsanin2017incremental,liu2016online}, have been introduced.
Tanahashi et al.~\cite{tanahashi2015efficient} extended the storyline generation algorithm for streaming data. 
To reduce both cognitive load and calculation cost, they utilized the previous steps' storylines to decide the new data points' layout. 
Crnovrsanin et al.~\cite{crnovrsanin2017incremental} developed the incremental graph layout based on $FM^3$~\cite{hachul2004drawing}---a fast force-directed layout algorithm.
To achieve faster calculation, they applied a GPU acceleration to $FM^3$.
Also, they designed the initialization, merging, and refinement steps of the graph layout to maintain the viewer's mental map. 
In addition, they used animation to provide smooth transitions from the previous to the current graph layout.
To support text stream analysis, Liu et al.~\cite{liu2016online} introduced a streaming tree cut algorithm to detect the incoming topics in time.
Also, their streamgraph visualization with a river metaphor can depict topics at different level-of-details to explore both global patterns from the accumulated results and local details from the new topics. 

Gansner et al.~\cite{gansner2013interactive} also worked on visualizing streaming text. 
They visualized topic relationships from the text data using a node-link diagram with a map metaphor~\cite{gansner2009gmap}, which can show clusters of texts clearly. 
To keep the viewer's mental map when updating the graph layout, they utilized MDS~\cite{torgerson1952} as a graph layout algorithm.
When calculating the new positions with MDS, their algorithm uses the previous nodes' positions as the initial positions to obtain a result that better maintains the mental map.  
Also, the algorithm applies the Procrustes transformation~\cite{borg1997modern} to reduce the positional changes caused by rotation and scaling between the successive MDS results.
Similarly, Cheng et al.~\cite{cheng2016framework} used MDS for showing an overview of similarities between temporal behaviors in streaming multivariate data. 
In addition, they introduced the concept of sliding MDS, which visualizes temporal changes in the similarities between selected points as line paths. 

The works of \cite{gansner2013interactive,cheng2016framework} are closely related to ours, in which we all utilize DR methods to visualize the relationships between the streaming data points. 
Both~\cite{gansner2013interactive} and~\cite{cheng2016framework} employed MDS as their DR methods. 
However, using MDS makes it difficult to incrementally update node positions based on new data points, as it requires a recalculation of all node positions every time a new data point appears (e.g., MDS needs several seconds to project 1,000 data points~\cite{yang2006fast}).
This scalability issue is particularly prominent when handling a large data size or if there is a frequent need to update the data. 
Our approach solves this scalability issue by using an incremental DR method.
We also take further steps to preserve the mental map by (1) minimizing the changes between current and incoming layouts and (2) using animation to smoothen the transition between the layouts.

\subsection{Dimensionality Reduction (DR) Methods}
\label{sec:relatedworkDR}
DR methods are essential tools in visualization that provide an overview of multidimensional data~\cite{liu2017visualizing,sacha2017visual}. 
For example, PCA~\cite{jolliffe1986principal} and MDS~\cite{torgerson1952} are popularly used in visualization research~\cite{sacha2017visual}.
The classical MDS, PCA, and many different variations of these two methods are categorized as linear DR methods~\cite{cunningham2015linear}.
In contrast, a non-linear DR method that is actively used in recent visualization studies is t-SNE~\cite{maaten2008visualizing,pezzotti2017approximated}. 
While a linear DR method is appropriate for showing the global structure of multidimensional data, a nonlinear DR method is useful to visualize the local structure of the data. 
One nonlinear DR method closely related to ours is the work by Goldberg and Ritov~\cite{goldberg2009local}.
While our method uses the Procrustes transformation~\cite{borg1997modern} to preserve the user's mental map, they used this transformation to find the low-dimensional representation which preserves the local structure of the multidimensional data.
A more comprehensive survey of DR methods can be found in ~\cite{van2009dimensionality,nonato2018multidimensional}.

As described in \autoref{sec:streamingVis}, one of the purposes of applying DR is to summarize time-series and/or multivariate data, including streaming data~\cite{cheng2016framework}.
For example, to identify anomalies from sensor networks, Steiger et al.~\cite{steiger2014visual} produced an overview of the sensors' behaviors. 
They used time-series similarity measures and then plotted the similarities with MDS. 
This method focuses on the comparison of each sensor's value over time.
In contrast, some visualizations calculate the similarity of the state of all data points at each time point, and then show their temporal differences. 
For example, Bach et al.~\cite{bach2016time} visualized the similarity of multivariate data between each time point by using MDS. 
van den Elzen et al.~\cite{van2016reducing} also applied similar methods. 
Rauber et al.~\cite{rauber2016visualizing} developed Dynamic t-SNE to compare the DR result for each time step. 
Dynamic t-SNE offers a controllable trade-off between how much temporal coherence is strictly kept and how much neighborhood relationships are precisely preserved in the t-SNE results.
J{\"a}ckle et al.~\cite{jackle2016temporal} introduced Temporal MDS Plots. 
They used $x$- and $y$-coordinates to represent time and MDS similarity, respectively. 
Also, they reduced the unnecessary rotation in the MDS results by flipping the $y$-coordinates based on their positions in the previous time point. 

Even though the stated existing works used DR methods for summarizing time-series data and addressed the issue of preserving a mental map, they still run into the issue of dealing with new data points due to the calculation cost.
This issue should be addressed for streaming data visualization. 
How to incorporate new data points to the existing result is one of the open problems in DR~\cite{strange2014open}.
Incremental DR methods have been developed to reduce the computation cost at each time point by updating the result incrementally. 
For example, methods like incremental PCA~\cite{oja1985stochastic,weng2003candid,ross2008incremental}, incremental Isomap~\cite{law2006incremental}, and incremental local linear embedding (LLE)~\cite{kouropteva2005incremental} are categorized as such.

In progressive visual analytics~\cite{muhlbacher2014opening,stolper2014progressive,turkay2018progressive}, researchers have started to apply incremental DR methods.
The main idea of progressive visual analytics is to provide useful intermediate results within a reasonable latency when the computational cost for an entire calculation is too high. 
Being able to produce usable results with a latency restriction is a common requirement for streaming data visualizations. 
For instance, Pezzotti et al.~\cite{pezzotti2017approximated} developed Approximated t-SNE (A-tSNE). 
Compared with t-SNE, A-tSNE stores the neighborhood information for each data point and only utilizes this information to refine the layout. 
Therefore, updating the layout in A-tSNE can work on each data point and its neighbors. 
By utilizing this characteristic, they also achieved an incremental update of the layout when adding or deleting points. 
While A-tSNE has been developed for progressive visual analytics, this is useful for streaming data visualization as well, as shown in their case study.
However, A-tSNE does not consider the mental-map preservation, and the added or deleted points would drastically affect the other data points' positions. 
Turkay et al.~\cite{turkay2017designing} used incremental PCA~\cite{ross2008incremental} in their system to generate an overview of multidimensional data within a second. 
They also employed an animated transition since the (incremental) PCA generates arbitrary rotations and flips in the plotted results at each iteration.
The animated transition acts a cognitive support that helps the user understand the incrementally updated PCA results. 

The two works~\cite{pezzotti2017approximated,turkay2017designing} are the most related works to ours. 
However, we approach a new problem where the streaming data has a different length of dimensions between each data point. 
In addition, when compared with the incremental PCA, as described in \cite{turkay2017designing}, we improve the incremental PCA in terms of preserving a mental map by using both position adjustment and animated transitions together. 

\section{Methodology}
\label{sec:methodology}
\begin{figure}[tb]
	\centering
	\captionsetup{farskip=0pt}
	\hspace*{-7pt}
    \subfloat[Without the geometric transformation]{
     \includegraphics[width=1.0\linewidth]{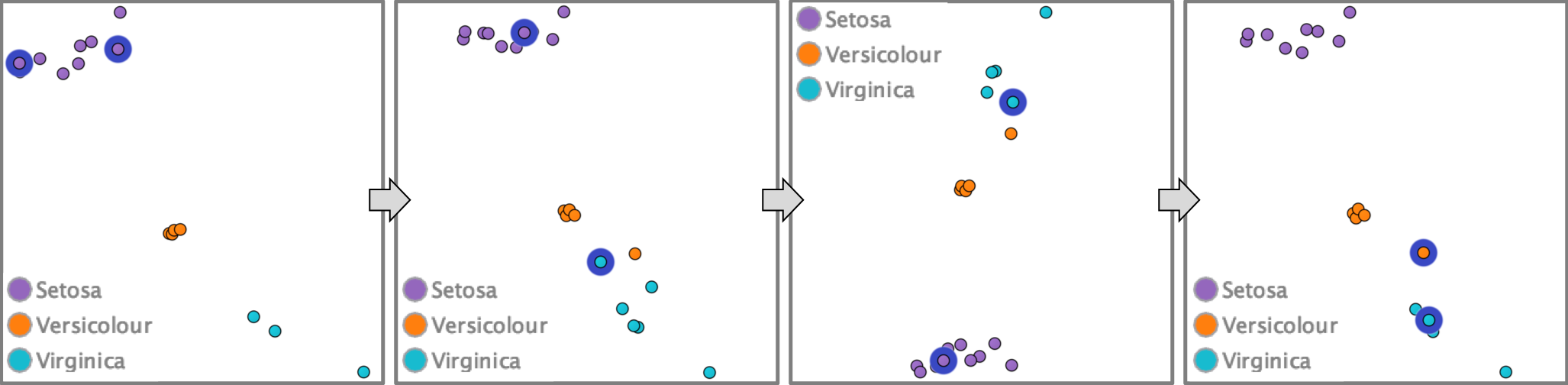}
     \label{fig:withoutGeomTrans}
    }
    \hspace*{-7pt}
    \\
    \hspace*{-7pt}
    \subfloat[With the geometric transformation]{
     \includegraphics[width=1.0\linewidth]{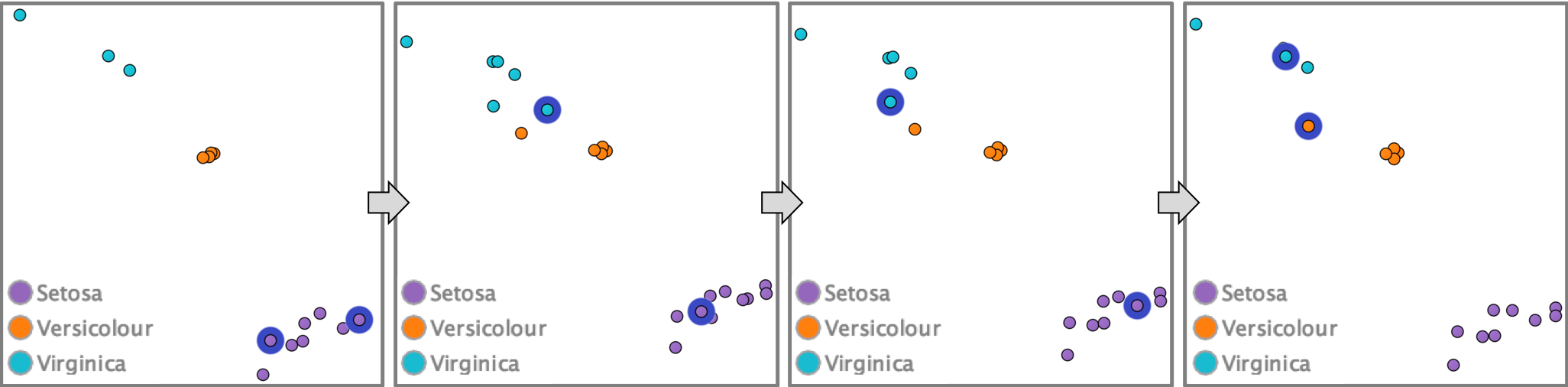}
     \label{fig:withGeomTrans}
    }
    \hspace*{-7pt}
    \caption{Comparison of the incremental PCA results for the Iris dataset (a) without and (b) with the geometric transformation. The point colors represent the Iris species. For each step, two points, highlighted in blue, are added to the result. In (a), a noticeable rotation and flipping can be seen. In (b), the plotted result is stable across all steps.
    }
	\label{fig:geomTrans}
\end{figure}

As mentioned, our goals are to effectively manage computational costs, preserve the viewer's mental map, and cope with data records with different numbers of dimensions.
To meet these goals, we made several design considerations for extending an existing incremental DR method. 
The resulting methodology is presented in this section. 
To better illustrate our work, we provide animations corresponding to  \autoref{fig:geomTrans}, \ref{fig:posEstimation}, and \ref{fig:autoTracking} online~\cite{supp}.
The source code for a major portion of our methods is also available in \cite{supp}.

\subsection{Incremental PCA}
\label{sec:incPCA}
Incremental DR methods incrementally update the lower-dimensional representations as new data points arrive~\cite{strange2014open}. 
Because the update only considers a small subset of the entire dataset, both computational complexity and memory usage can be reduced. 
We employ incremental PCA~\cite{oja1985stochastic,weng2003candid,ross2008incremental} because of PCA's popularity in the visualization community~\cite{sacha2017visual}.

Among incremental PCA algorithms, we choose the model by Ross et al.~\cite{ross2008incremental}---an extension of the Sequential Karhunen-Loeve algorithm~\cite{levey2000sequential}. 
To apply the model, several parameters need to be pre-determined: $d$, the number of dimensions that a data point has, $n$, the number of data points processed so far, $m$, the number of accumulated new data points for the next update (the model requires $m \geq 2$), and $k$, the number of principal components to use. 
One of the main advantages of utilizing this model is its relatively low computation and space complexity. 
The time and space complexity of an ordinary PCA~\cite{jolliffe1986principal} are $O(d^2(n+m) + d^3)$ and $O(d^2)$, respectively. 
In contrast, the model by Ross et al. reduces the time and space complexity to  $O(dm^2)$ and $O(d(k+m))$, respectively.
This is because, based on only the partial singular value decomposition (SVD) with $m$ new data points, the model incrementally updates the SVD for all data points, which is required for PCA.
Because we usually have a fairly small $m$ value in streaming data applications, the computational cost can be scaled down significantly. 

There are other benefits in applying Ross et al.'s~model. 
Unlike other incremental PCA methods (e.g., \cite{levey2000sequential,hall1998incremental}), this model constantly updates the sample mean, which is subsequently used for updating the eigenbasis of PCA. 
As a result, utilizing the model does not require setting up a learning phase, which addresses two common issues in handling streaming data: (1) we do not need to wait until a certain amount of data is accumulated to perform an update; (2) we always have an updated sample mean for incorporating new incoming data. 

Furthermore, in the model, we can set a ``forgetting factor'', denoted $f$, which provides a way to reduce the contributions of past observations (existing data points) to the latest result. 
The value of $f$ ranges from $0$ to $1$, where $f=1$ means no past results will be forgotten. 
Whereas, when $f < 1$, the contributions of past observations are gradually decreased as new data points are obtained. 
The effective size of the observation history (the number of observations which will affect the PCA result) equals to $m/(1-f)$~\cite{ross2008incremental}.
For example, when $f=0.998$ and $m=2$, only the most recent $1,000$ observations are effective. 
By utilizing $f$, we can support both incremental addition of new data points and incremental deletion of past observations.
Once the number of observations reaches the effective size, the effects of the past observations to the PCA calculation are ignored. 
As a result, we can choose to either keep or delete the past observations based on the user's need. 
\subsection{Preserving the Viewer's Mental Map}
\label{sec:mentalMap}

The results directly derived from the incremental PCA would have an arbitrary rotation and/or flipping of data points at subsequent time steps.
\autoref{fig:withoutGeomTrans} shows an example demonstrating this issue using the Iris dataset~\cite{fisher1936use,anderson1936species}.
If this issue is not handled properly, it is difficult for the viewer to follow the updates in the visualization as the mental map can easily get lost during the analysis. 
Our solution is to minimize the moving distance of the same set of data points between two subsequent time steps by applying a geometric transformation and then using animations for smoother transitions. 

The PCA's flipping issue is known as the ``sign ambiguity'' problem and some possible solutions for visualizations have been proposed by Bro et al.~\cite{bro2008resolving}, Jeong et al.~\cite{jeong2009understanding}, and Turkay et al.~\cite{turkay2017designing}. 
However, these methods do not consider the issue of arbitrary rotation of data points.
To address both the flipping and arbitrary rotation, we apply the Procrustes transformation~\cite{schonemann1970fitting,akca2003generalized,gower2004procrustes}.
The Procrustes transformation is used to find the best overlap between two sets of positions (i.e., the previous and current PCA results in our case) by using only translation, uniform scaling, rotation, reflection, or a combination of these transformations.
The objective function to find the geometric transformation for the best overlap can be written as:
    \begin{equation}
        \label{eq:transform}
        \mathrm{Minimize\ } \|c (\mathbf{P'} + \mathbf{v}\boldsymbol{\tau}^\mathsf{T}) \mathbf{R} - \mathbf{P}\|^2
    \end{equation}
where $\mathbf{P}$ and $\mathbf{P'}$ are ($n\times k$) matrices that contain the first $k$ principal component values of $n$ data points for the previous and current PCA results, respectively.
$n$ is the number of data points found in both the previous and current PCA results.
$\boldsymbol{\tau}$ is a ($k \times 1$) vector which translates data points of $\mathbf{P'}$ with $\mathbf{v}$, while $\smash{\mathbf{v} = (1\ 1\ \ldots 1)^\mathsf{T}}$ is a ($n \times 1$) vector. 
$c$ represents the uniform scale factor ($c \in \mathbb{R}$). 
$\mathbf{R}$ is a ($k \times k$) orthogonal rotation matrix, which handles rotation and reflection. 

The Procrustes transformation starts by translating $\mathbf{P'}$ so that the centroid of $\mathbf{P'}$ is placed at the centroid of $\mathbf{P}$.
Let $\bar{\mathbf{p}}$ and $\bar{\mathbf{p}}'$ be ($k \times 1$) vectors which represent the centroids of $\mathbf{P}$ and $\mathbf{P'}$, respectively.
We can compute the translation vector $\boldsymbol{\tau} = \bar{\mathbf{p}} - \bar{\mathbf{p}}'$.
Now, ($\smash{\mathbf{P'} + \mathbf{v}\boldsymbol{\tau}^\mathsf{T}}$) represents the translated result. 
The next step is scaling ($\smash{\mathbf{P'} + \mathbf{v}\boldsymbol{\tau}^\mathsf{T}}$) to eliminate the scaling differences from $\mathbf{P}$. 
This can be achieved by matching the root mean square distances of $\mathbf{P}$ and ($\smash{\mathbf{P'} + \mathbf{v}\boldsymbol{\tau}^\mathsf{T}}$) from the centroid of $\mathbf{P}$.
This scaling factor $c$ can be calculated as $\smash{c = \| \mathbf{P} - \mathbf{v} \bar{\mathbf{p}}^\mathsf{T}\| / \| \mathbf{P'} - \mathbf{v} \bar{\mathbf{p}}^{'\mathsf{T}}\|}$.
Lastly, the Procrustes transformation computes $\mathbf{R}$ for optimal rotation and reflection.
To obtain $\mathbf{R}$, singular-value decomposition (SVD) is performed on $\smash{c \mathbf{P}^\mathsf{T} (\mathbf{P'} + \mathbf{v}\boldsymbol{\tau}^\mathsf{T})}$ (i.e., $\smash{c \mathbf{P}^\mathsf{T} (\mathbf{P'} + \mathbf{v}\boldsymbol{\tau}^\mathsf{T}) = \mathbf{U}\Sigma\mathbf{V}^\mathsf{T}}$).
Then, $\smash{\mathbf{R} = \mathbf{V}\mathbf{U}^\mathsf{T}}$.
Please refer to \cite{schonemann1970fitting,akca2003generalized,gower2004procrustes} for more information on why this $\mathbf{R}$ provides the optimal rotation and reflection.

Now with $c$, $\boldsymbol{\tau}$, and $\mathbf{R}$, we can transform the data points in the current PCA result to minimize their moving distance from the previous result. 
\autoref{fig:geomTrans} shows a comparison between results with and without applying the transformation. 
We can see that the transformation reduces unnecessary changes across the time points. 
Note that the time complexity of the Procrustes transformation is $O(k^2 n + k^3)$.
For visualization purpose, usually $k\leq3$, and thus, this transformation is fast enough to handle streaming data. 

Furthermore, we animate the change of the data points to maintain the coherence between each subsequent step.
We utilize the staged transitions from Bach et al.~\cite{bach2014graphdiaries}, which was originally developed for visualizing dynamic node-link diagrams.
The transitions consist of three stages: (1) fading-out the data points that need to be removed; (2) moving the remaining data points from their previous positions to their new positions; (3) fading-in the new incoming data points. 

\begin{figure}[tb]
    \captionsetup{farskip=0pt}
	\centering
	\hspace*{-7pt}
        \includegraphics[width=1.0\linewidth]{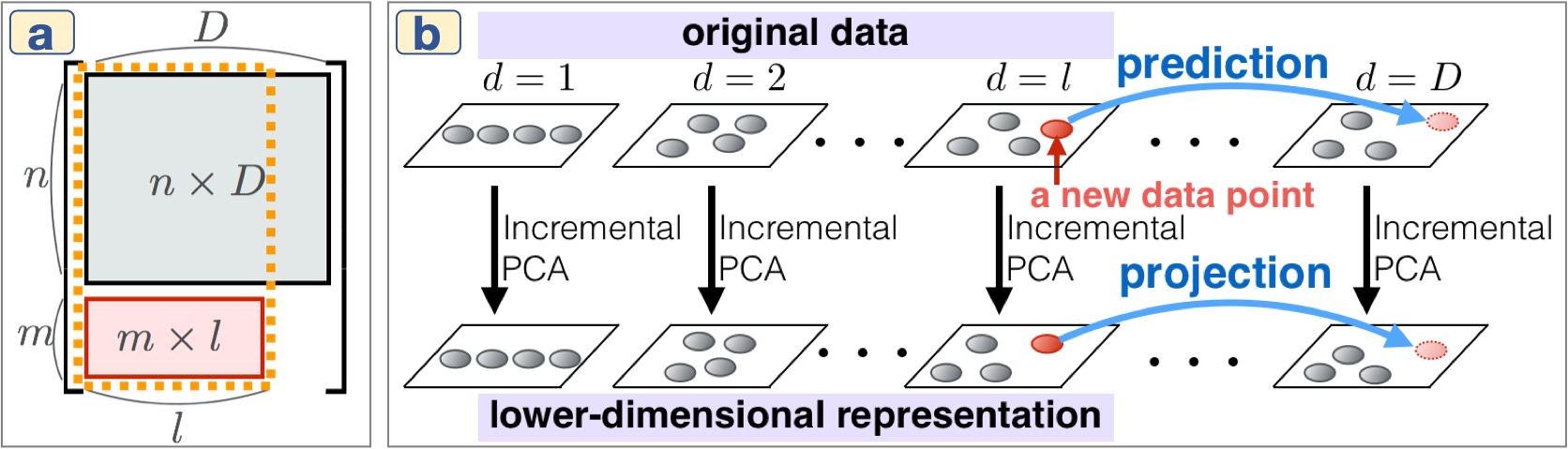}
    \hspace*{-7pt}
   	\caption{A relationship between variant dimensions and the incremental PCA results. (a) shows the data shape. While $n$ stored data points have $d=D$ (the gray area), $m$ new data points have $l$ dimensions (the red area). (b) shows the relationships between the data and its lower-dimensional representation generated by the incremental PCA. Gray and red points represent the stored and new data points, respectively. The PCA can be applied to only the grey area or the area within the orange border in (a). Thus, we need to apply a prediction or projection to obtain the lower-dimensional representation for all ($n+m$) data points where $d=D$.}
	\label{fig:variantDim}
\end{figure}
\subsection{Position Estimation for Dealing with a Non-uniform Number of Dimensions}
\label{sec:varDim}

When the streaming data contains data points with a non-uniform number of dimensions, ordinary incremental PCA cannot be directly used. 
We first describe the problem of applying incremental PCA for such a case. 
Then, we present an algorithm that addresses the issue. 

Let $D$ be the complete number of dimensions (or features) that each data point can contain. 
From the data stream, $n$ past data points have already gathered the information from all dimensions ($d=D$).
On the other hand, some $m$ new data points could have an incomplete $l$ number of dimensions, ranging from anywhere between $d=1$ and $d=D$. 
Consider the following example: in an online transaction stream, if we assume that there are $D$ steps to reach to the purchase checkout, we have stored the history of $n$ users' time spent at each step. 
However, $m$ new users just finished the $l$-th step ($1 \leq l \leq D$) and, thus, we only have access to their time information for the first $l$ steps.

If we want to compare the $m$ new data points to the $n$ existing data points using a DR method, one common method is to fill in the unknown values with a derived value (e.g., the mean or median value from the $n$ data points). Another alternative is to apply DR only to the first $l$ dimensions. 
Each approach has its limitation in streaming data applications. 
The first method does not capture the characteristics of the data well (e.g., correlations)~\cite{dray2015principal}, while the second method requires a re-calculation of the PCA every time the value of $l$ changes. 

\autoref{fig:variantDim} shows the relationship between the data points and the results after applying the incremental PCA for each different number of dimensions.
When $m$ new points have $l$ dimensions, we can obtain the PCA results up to the $l$ dimensions. 
Because $l \leq D$, we can only apply PCA to the $(n+m)$ data points using $l$ dimensions (the area within the orange outline in \autoref{fig:variantDim}a). Alternatively, if we want to apply PCA using the full dimension (i.e., $d=D$), we can only do so with the $n$ data points (the gray area in \autoref{fig:variantDim}a). 

There are two possible solutions for employing PCA to obtain the lower-dimensional representation for $D$ dimensions with the $(n+m)$ data points, as indicated with the blue arrows in \autoref{fig:variantDim}b. 
The first method is to predict the values for the rest of the dimensions by using some machine learning or estimation methods~\cite{loisel2018comparisons} (e.g., linear regression). 
Then, we can apply incremental PCA to all ($n+m$) data points.
The second method is to project the $m$ data points' positions from the PCA result of $d=l$ onto the PCA result of $d=D$. 
Compared to the first method, the second method executes in a simpler manner as it does not require choosing a proper model for a specific dataset, tuning the model used for a prediction-based method, or having an excessive computational cost.

We, therefore, use the second method and introduce a position estimation method.
This method estimates where the positions of the new data points would be in the PCA result of $d=D$ by utilizing the distances between the new data points and the existing points (which already have the full dimension information $d=D$) in the PCA result of $d=l$. 
The estimation method proceeds in the following manner: 
first, we apply the incremental PCA for $d=l$; then, we project the positions of $m$ new data points to the PCA result of $d=D$, such that it maximally preserves the distance relationships between the new and existing data points in the PCA result of $d=l$.
This idea is based on the assumption that a new data point will likely have a similar relationship with the other data points in the remaining dimensions. 
The objective function for this optimization problem can be written as:
    \begin{equation}
        \label{eq:distErrMin}
        \operatorname*{argmin}_\theta \sum_{i=1}^{n}\left(s_{ui} - \alpha s'_{ui} \right)^2 = \operatorname*{argmin}_\theta \sum_{i=1}^{n}\left(s_{ui} - \alpha \left\lVert \mathbf{x} - \mathbf{q}_i \right\rVert \right)^2
    \end{equation}
where $\theta$ consists of the parameters of $\alpha$ and $\mathbf{x}$ ($\alpha \in \mathbb{R}$, $\mathbf{x} \in \mathbb{R}^2$). 
$s_{ui}$ and $s'_{ui}$ are the distances from a new data point $u$ to the $i$-th existing data point in the PCA results of $l$ and $D$ dimensions, respectively.
$\mathbf{q}_i$ is the position of the $i$-th existing data point in the PCA result of $d=D$. 
$\mathbf{x}$ represents the estimated position of the new data point $u$ in the PCA result of $d=D$ using this objective function. 
$\alpha$ is used for eliminating the scaling difference between each PCA result.
The idea of adding data points to the DR result based on the distance relationships with the existing data points is similar to pivot-based MDS algorithms~\cite{morrison2003fast,jourdan2004multiscale} which target on reducing the computational cost.

We apply a gradient descent algorithm~\cite{ruder2016overview} to find the parameters $\theta$ in \autoref{eq:distErrMin}. 
Specifically, we use Adadelta~\cite{zeiler2012adadelta}, as this model can automatically adapt the learning rate for each parameter without providing a default value~\cite{ruder2016overview}.
After obtaining $\theta$, we place the new data point $u$ at the position $\mathbf{x}$ in the PCA result of $d=D$.
Since there are $m$ new data points, we apply this calculation for each new point.
Note that $\alpha$ may be a different value for each new point.
We chose to apply \autoref{eq:distErrMin} to each new point separately rather than finding the best common $\alpha$ for all new points, as the latter requires much more computations.

\begin{figure}[tb]
    \captionsetup{farskip=0pt}
    	\centering
	\hspace*{-9pt}
    \subfloat[Without new points]{
     \includegraphics[width=0.325\linewidth,height=0.31\linewidth]{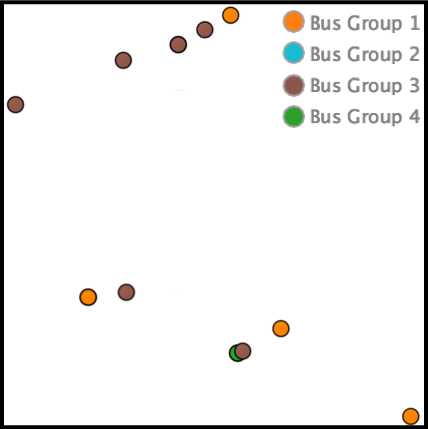}
     \label{fig:posEstimation_a}
    }
    \hspace*{-6pt}
    \subfloat[\!Estimation with $d\!=\!1$]{
     \includegraphics[width=0.325\linewidth,height=0.31\linewidth]{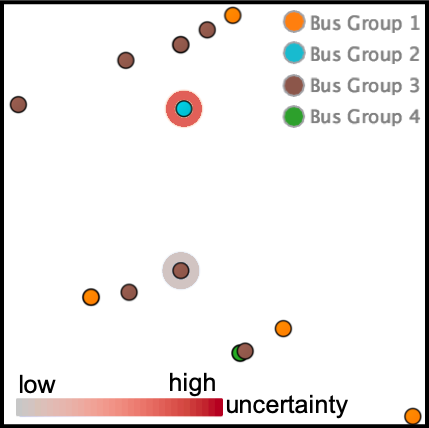}
     \label{fig:posEstimation_b}
    }
    \hspace*{-6pt}
    \subfloat[\!Estimation with $d\!=\!2$]{
     \includegraphics[width=0.34\linewidth,height=0.31\linewidth]{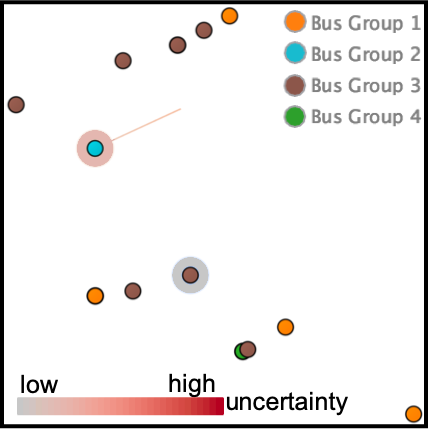}
     \label{fig:posEstimation_c}
    }
    \hspace*{-7pt}
    \\
    \hspace*{-9pt}
    \subfloat[\!Estimation with $d\!=\!3$]{
     \includegraphics[width=0.325\linewidth,height=0.31\linewidth]{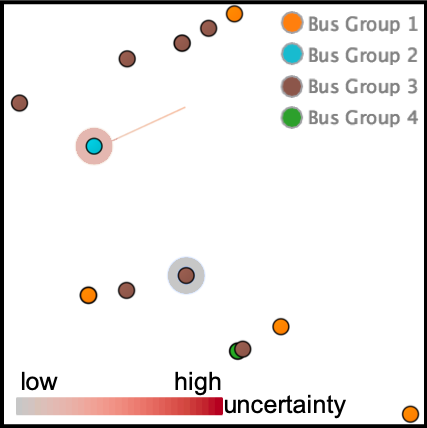}
     \label{fig:posEstimation_d}
    }
    \hspace*{-6pt}
    \subfloat[\!Estimation with $d\!=\!4$]{
     \includegraphics[width=0.325\linewidth,height=0.31\linewidth]{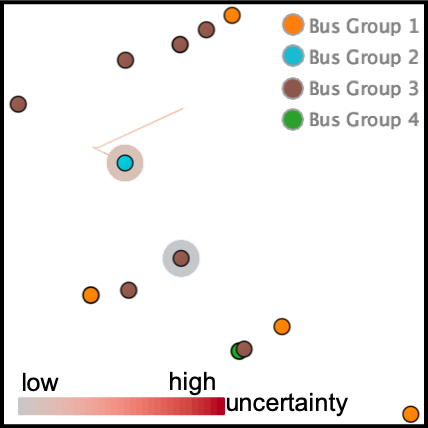}
     \label{fig:posEstimation_e}
    }
    \hspace*{-6pt}
    \subfloat[PCA\,result\,with\,new\,points]{
     \includegraphics[width=0.34\linewidth,height=0.31\linewidth]{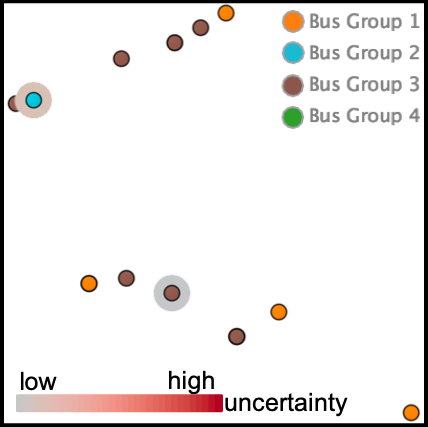}
     \label{fig:posEstimation_f}
    }
    \hspace*{-7pt}
    \caption{Visualizations with the position estimation method. The bus transportation dataset from \cite{transitfeeds} is used for this visualization. Each point represents one bus and each color represents a bus group defined by ``Block ID'' in \cite{transitfeeds}.
    The time duration between each bus stop is used as a value for each dimension. Each bus passes through five stops and has four time durations ($D=4$). The position transitions of two new buses are shown from (a) to (f). In (a), the PCA result of $d=D$ are shown without plotting the new buses. From (b) to (e), the two buses are plotted with the position estimation method where $d=1$ to $d=D$. Then, the PCA result is updated in (f). The outer-ring color represents the uncertainty as described in \autoref{sec:uncert}.
    From (c) to (e), paths of the two buses are also visualized with the corresponding uncertainty colors. We can see that a bus with higher uncertainty in the top-left moves farther away from the final result in (f).
   	}
	\label{fig:posEstimation}
\end{figure}

Once the new data points obtain the values of the additional dimensions (e.g., changing from $l$ to $l+1$), the positions of the new data points will be updated by applying this method incrementally. 
\autoref{fig:posEstimation} shows an example of the ongoing updates from the position estimation results.
Same as \autoref{sec:mentalMap}, we show the transitions of the new points' positions with the staged transitions.  

\begin{figure}[tb]
    \captionsetup{farskip=0pt}
	\centering
    \includegraphics[width=1.0\linewidth,height=0.28\linewidth]{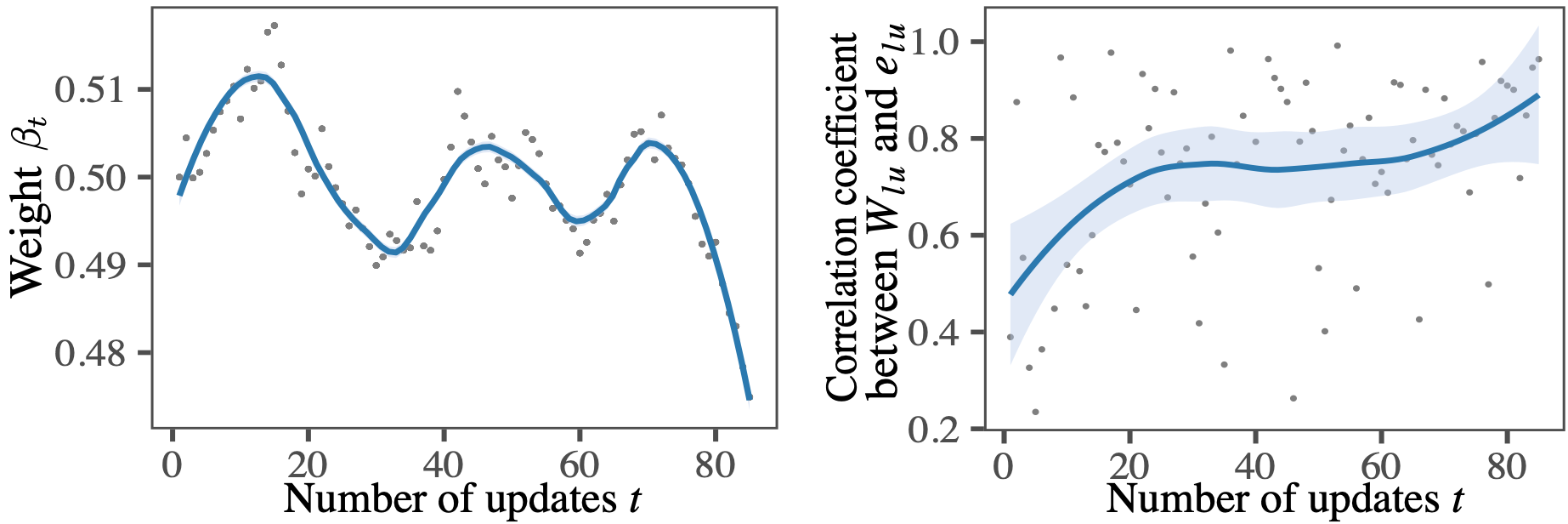}
    \caption{Results of the automatic selection of $\beta$. The same dataset as \autoref{fig:posEstimation} is used. The left shows a scatterplot of the number of updates $t$ and the weight $\beta_t$. The right is a scatterplot of the number of updates $t$ and the Pearson correlation coefficient between the uncertainty ${W_l}_u$ and the error ${e_l}_u$. The blue lines show the smoothed lines with LOESS~\cite{cleveland1979robust}, while the light blue areas represent the 95\% confidence intervals.} 
	\label{fig:uncertAnalysis}
\end{figure}

\subsection{Visualizing Uncertainty of the Position Estimation}
\label{sec:uncert}

Our position estimation method introduces two uncertainties. 
Both uncertainties represent how inaccurately the new point is projected onto the PCA result of $d=D$. 
A data point with higher uncertainty has a higher chance of moving drastically until its position is updated again with the next incremental PCA calculation (when $d=D$). 
 
The first uncertainty is derived from the optimization using \autoref{eq:distErrMin}. 
The cost remaining after the optimization can indicate how the distance between each pair of data points in the PCA result of $d=l$ is different from the one derived from the PCA result of $d=D$.
We calculate this uncertainty in a range from $0$ to $1$. 
Similar with the ``strain'' in the classical MDS~\cite{torgerson1952}, the uncertainty ${U_l}_u$ ($0 \leq {U_l}_u \leq 1$) for the new data point $u$ with $l$ dimensions can be calculated with: 
    \begin{equation}
        \label{eq:uncert1}
        {U_l}_u = \left(
                    \frac{\sum_{i=1}^{n}\left(s_{ui} - \alpha s'_{ui} \right)^2}
                    {\sum_{i=1}^{n} s_{ui}^2} 
                \right)^{1/2}
    \end{equation}

The second uncertainty comes from the fact that a new data point does not have the values for all of the $D$ dimensions when the position estimation method is applied (the new point has only $l$ dimensions).
We utilize the principal component (PC) loading derived from the PCA to calculate this uncertainty.
The PC loading represents the correlation between the original variables and the PCs. 
This can indicate how much variance each dimension contributes to each PC. 
The PC loading ($w_{ij}$) of $j$-th dimension to the $i$-th PC can be written down as: $w_{ij} = \smash{\sqrt{\lambda_i}h_{ij}}$ where $\lambda_i$ is the eigenvalue for the $i$-th PC and $h_{ij}$ is the $j$-th element of the eigenvector $\mathbf{h}_i$ which corresponds to $\lambda_i$. 

By using $w_{ij}$, the uncertainty $V_l$ ($0 \leq {V_l} \leq 1$) for the new data points with $l$ dimensions can be written down as:
    \begin{equation}
        \label{eq:uncert2}
        V_l = 1 - \frac{1}{k} \sum_{i=1}^{k}\left(\frac{  \sum_{j=1}^{j=l}|w_{ij}| } { \sum_{j=1}^{j=D}|w_{ij}| } \right)
    \end{equation}
Here, $\smash{\sum_{j=1}^{j=l}|w_{ij}|/\sum_{j=1}^{j=D}|w_{ij}|}$ is the proportion of the sum of PC loading when we have $l$ dimensions to the sum of PC loading for all dimensions. 
This means how much information of the $i$-th PC is already covered when we have $l$ dimensions.
By taking the average of these proportions for all the PCs that are utilized in the visualization (the first and second PCs when the result is in 2D), we can obtain a percentage that describes how much of the visualized information of the data the PCA result of $l$ dimensions explains.
Therefore, by subtracting this value from 1, $V_l$ can indicate how much information has not been considered during the position estimation process.
Note that $V_l$ remains the same for all $m$ new points, while ${U_l}_u$ is different for each new data point.

To account for both uncertainties, we can compute a combined uncertainty ${W_l}_u$ with ${W_l}_u = \beta {U_l}_u + (1 - \beta) V_l$ for each new data point $u$. 
The value of $\beta$ ($0 \leq \beta \leq 1$) serves as a parameter for controlling the weight for either uncertainty. 
$\beta$ can be defined manually or determined automatically (see the description in the following paragraph). 
We encode the combined uncertainty ${W_l}_u$ with an outer-ring color for each plotted point using a red sequential colormap, as shown in \autoref{fig:posEstimation}. 
The saturated red outer-ring refers to a high uncertainty value.
In addition, we draw a path for each new data point's movement with gradient colors to represent the uncertainties at the corresponding source and target positions. 
This allows us to see the change of the data positions and uncertainties.

Selecting a proper value of $\beta$ is not trivial because the user may not have a clear criterion to follow.
Thus, we provide an automatic method to help users decide the value of $\beta$. 
Let $\sigma_{ui}$ be the distance between a new data point $u$ and an existing point $i$ in the updated PCA result after $u$ reaches $d=D$. 
The mean absolute error ${e_l}_u$ for the estimated distance relationship of $u$ when $u$ has $l$ dimensions of the data is:
    \begin{equation}
        \label{eq:error}
        {e_l}_u = \frac{1}{n} \sum_{i=1}^{n} \left| \sigma_{ui} - s'_{ui}\right|
    \end{equation}
${W_l}_u$ should be an indication of this future error ${e_l}_u$. 
Thus, we can assume that ${e_l}_u$ is proportional to ${W_l}_u$ (i.e., ${e_l}_u \propto {W_l}_u$). 
We calculate a proper $\beta$, as $\beta$ adjusts the balance between $U_{l_u}$ and $V_l$, to obtain this proportional relationship.
From ${e_l}_u \propto {W_l}_u$, we obtain ${e_l}_u = \rho {U_l}_u + \phi V_l$ where $\rho = a \beta$ and $\phi= a(1 - \beta)$ ($a \in \mathbb{R}$). 
Since $\beta$ can be calculated with $\beta = \rho / (\rho + \phi)$, we want to obtain $\rho$ and $\phi$.

First, we utilize the fact that the estimated positions when $l=D$ have no uncertainty for $V_l$ (i.e., $V_D = 0$).
Then, we can consider that ${e_D}_u$ is proportional to ${U_D}_u$ (i.e., ${e_D}_u \propto {U_D}_u$).
Second, for all dimensions $1 \leq l \leq D$, we approximate ${U_l}_u$ with $E[U]_u = \sum_{l=1}^{D}{U_l}_u / D$ (i.e., ${U_l}_u \simeq E[U]_u \forall l$). 
We then obtain a relationship of ${e_l}_u - {e_D}_u \propto  V_l$.
Then, we can approximate $\beta$ with the following equations:
    \begin{equation}
        \rho = \frac{ {e_D}_u }{ E[U]_u }, \ \ 
        \phi = \frac{ \sum_{i=1}^{D}({e_i}_u - {e_D}_u) }{  \sum_{i=1}^{D}V_i }
    \end{equation}
    \begin{equation}
        \beta = \frac{\rho}{\rho + \phi}
    \end{equation}

However, this $\beta$ cannot be calculated when we estimate the position for the new point $u$ when we have only $l$ dimensions since the value of $\sigma_{ui}$ in \autoref{eq:error} is still unknown at that time. 
Thus, we obtain the approximated $\beta$ by calculating the gradient of $\beta$ from the previous PCA result. 
Let $\beta_t$ be the $\beta$ for the PCA result after updating $t$ times (i.e., after applying process a3 in \autoref{fig:flowchart} $t$ times). 
By using the same update method as Adadelta~\cite{zeiler2012adadelta}, $\beta_{t+1}$ can be estimated from $\beta_t$ with:
    \begin{equation}
        g_t = \beta_t - \frac{\rho}{\rho + \phi}, \ \ 
        \Delta \beta_t = - \frac{RMS[\Delta \beta]_{t-1}}{RMS[g]_t}g_t
    \end{equation}
    \begin{equation}
        \beta_{t+1} = \beta_t + \Delta \beta_t
    \end{equation}
where $RMS[\cdot]$ is the root mean square and $\beta_0 = c$ is set as the initial parameter for $\beta$. 
By default, we set $c = 0.5$. 
This method automatically adjusts the weight $\beta$, as the PCA result is updated. 
As a result, the user can keep observing the uncertainty with well-balanced weights. 

\autoref{fig:uncertAnalysis}(left) shows an example of the automatic selected $\beta$. 
We used the same dataset as \autoref{fig:posEstimation} (the bus transportation dataset from \cite{transitfeeds}).
We observe that $\beta$ keeps increasing to more than $0.5$ when $t < 10$ (i.e., the uncertainty ${U_l}_u$ has more influence to the error ${e_l}_u$), while we see the inverse relationship when $t \geq 70$ (i.e., $\beta$ keeps decreasing from $0.5$ implying that the uncertainty $V_l$ has more influence to ${e_l}_u$).
\autoref{fig:uncertAnalysis}(right) shows the transition of the Pearson correlation coefficient (PCC) between the combined uncertainty ${W_l}_u$ and the error ${e_l}_u$ (more specifically, sets of ${W_l}_u$ and ${e_l}_u$ with $l\!\!=\!\!\{1,\cdots,D\}$ and $u\!\!=\!\!\{1,\cdots,m\}$ for each $t$).
First, we can see that ${W_l}_u$ and ${e_l}_u$ have a positive association at each time point. 
Therefore, ${W_l}_u$ can well represent the uncertainty of the placement of each new data point.
Also, the increase of PCC can be seen when $t < 20$.
This indicates that the automatic update of $\beta$ contributes to better obtainment of the uncertainty ${W_l}_u$.

\subsection{Automatic Tracking}
\label{sec:autoTracking}
\begin{figure}[tb]
    \captionsetup{farskip=0pt}
	\centering
     \includegraphics[width=1.0\linewidth]{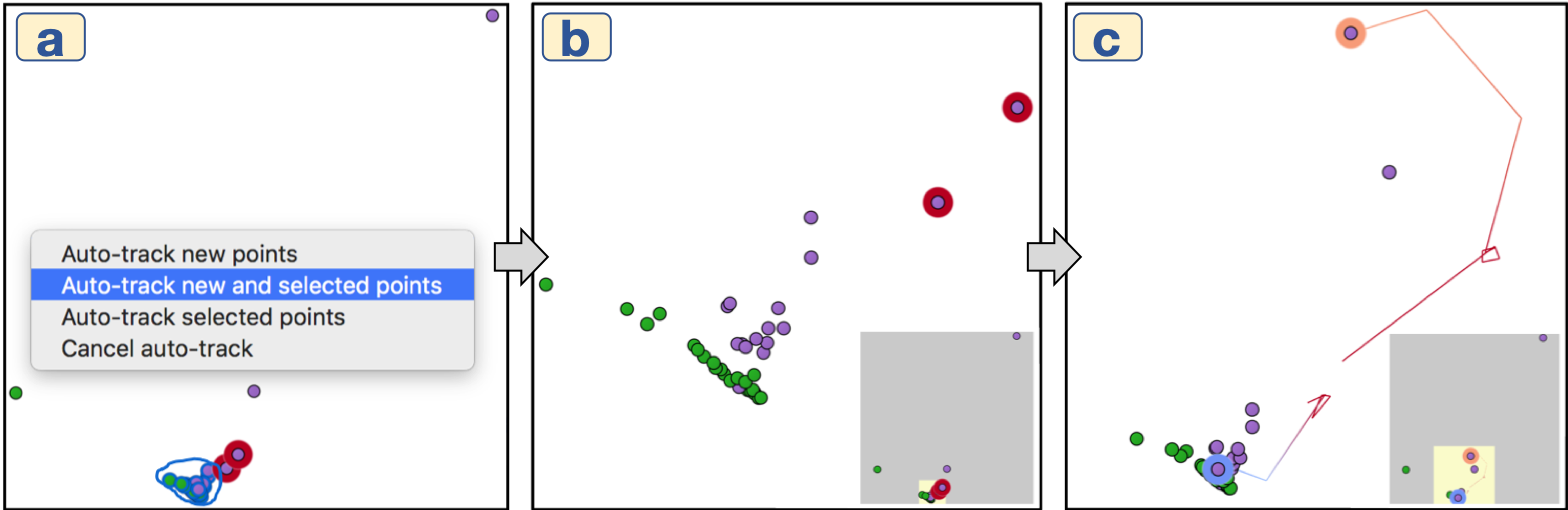}
    \caption{An example of the automatic tracking. In (a), since the outliers make the result sparse, we select points in the dense area and choose to track the selected points and new points from the dialog menu. Then, zooming and panning are automatically applied to focus on the tracking points in (b). When the positions of the tracking points are updated, the focus area is also automatically updated, as shown in (c). 
    }
	\label{fig:autoTracking}
\end{figure}

In \autoref{sec:mentalMap}, we described a method that helps the user follow the frequent changes that will occur in streaming data visualizations. 
There are two additional considerations that need to be taken into account. 
One is that the estimated position calculated from \autoref{sec:varDim} can be outside of the range~\cite{krstajic2013visualization} of the PCA result. 
In this case, to avoid failing to inform the user of an important change, the visualization should update its ranges of axes or have an indicator to notify the user that there are points outside of the ranges. 
The second consideration relates to outliers when using linear DR methods, including PCA.
When the data includes an outlier, DR methods will project the outlier to a position which is far away from other data points. For example, in \autoref{fig:autoTracking}a, a purple point at the top-right and a green point at the bottom-left are two outliers. 
In this case, the user may be interested in only keeping track of the data points that are not outliers. 

To address these issues, we provide an automatic tracking mechanism that allows the user to stay focused on the data points of his/her interest. 
\autoref{fig:autoTracking} shows the process of the automatic tracking.
First, the user indicates the data points of interest through some selection method, such as a lasso selection.
For example, in \autoref{fig:autoTracking}a, the user chooses the data points by lassoing and then selects the selected data points and incoming new data points as tracking targets from a dialog menu. 
Next, zooming and panning are applied to show the selected points in the center of the scaled window, as shown in \autoref{fig:autoTracking}b. 
Zooming and panning will be applied again when the plotted result is updated by either new estimated positions, the addition of new points, or a recalculation of the PCA result (\autoref{fig:autoTracking}c). 

However, when a large change occurs, it is difficult for a user to preserve his/her mental map. 
To help maintain the mental map, we use animated transitions for zooming and panning (referred to as the view-level transitions) in addition to the three staged animated transitions~\cite{bach2014graphdiaries} (referred to as the visual-structure level transitions).

We apply the animations in the following order: panning, zooming in, removing, moving, and adding data points for the cases of zoom-in animations. For the zoom-out animations, we follow the order of zooming out, panning, removing, moving, and adding data points. 
We have tested multiple alternative designs.
First, we used the visual-structure level transitions and the view-level transitions in parallel. 
However, this caused many changes to happen simultaneously and the result was hard to follow. 
Another option was to use the visual-structure level transitions before the view-level transitions. 
In this case, actions, such as removing, moving, or adding data points, could happen outside of the axes ranges. 
As a result, there was a potential issue of failing to inform the changes to the user. 
The last consideration was the order of zoom and pan in the view-level transitions. 
When we first animated zoom and then pan, the visualization was zoomed into the unrelated area of the selected points.
This also made it difficult to follow the changes.
A similar result occurred when we first animated pan then zoom-out. 
This issue also happened when applying zooming and panning in parallel, similar to \cite{van2003smooth}.
Thus, we decided to employ different orders of steps based on whether it required zooming-in or zooming-out.

In addition, we provide a mini-map to help the user grasp which part of the plot he/she is looking at after panning and zooming.
An example of visualization with animated transitions can be found online~\cite{supp}.

\section{Performance Evaluation}

We demonstrate that our methods are fast enough for handling streaming data through an evaluation of computational performance for each method.
As an experimental platform, iMac (Retina 5K, 27-inch, Late 2014) was used. 
It has 4 GHz Intel Core i7, 16 GB 1,600 MHz DDR3.

\begin{figure}[tb]
	\centering
	\includegraphics[width=0.85\linewidth]{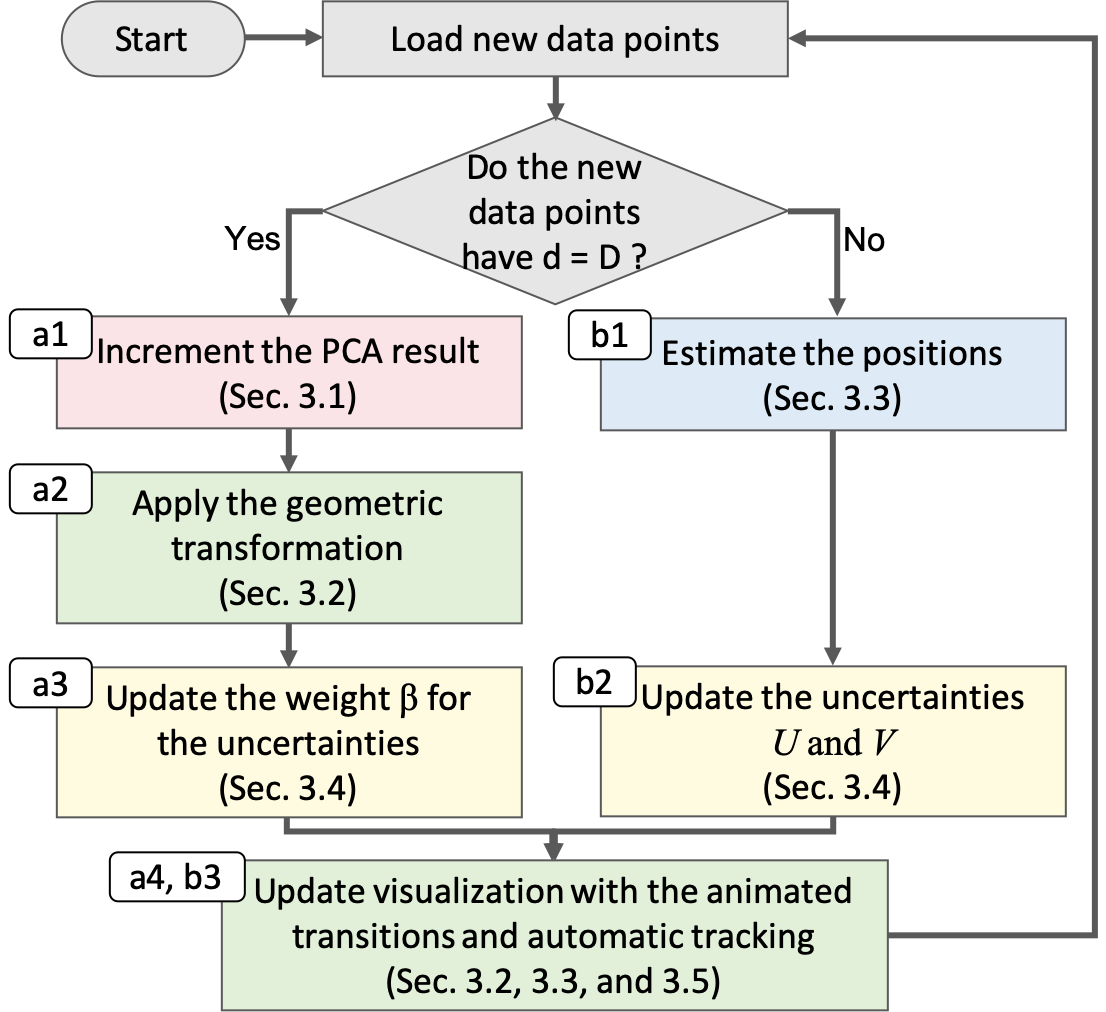}
   	\caption{A flowchart of the streaming data visualization using incremental PCA. The red, green, blue, and yellow are process blocks that correspond to the methods for dealing with the computational cost, the viewer's mental map, and the non-uniform number of dimensions, the uncertainty visualization, respectively. 
   	}
	\label{fig:flowchart}
\end{figure}

\autoref{fig:flowchart} shows the flowchart of the overall process starting from receiving the new data points to visualizing the results.
There are two main flows on how to deal with new data points based on whether they have the values for all $D$ dimensions (processes a1--a4 in \autoref{fig:flowchart}) or do not (b1--b3). 
Since the completion time of a4 and b3 mainly depends on the duration of animated transitions, we only measure the completion times of the other processes. 
To run the experiment, we generate datasets containing data points with a different number of dimensions ($D$=10,\,100,\,1,000), and all values are randomly assigned in the $[-1:1]$ range. 
In addition, we set the frequency for updating the visualized results at every $2$ new data points ($m=2$). 
The geometric transformation and position estimation method are applied for 2D points. 
The maximum number of iterations for running the Adadelta optimization (\autoref{eq:distErrMin}) is bounded at 1,000. 

\begin{table}[tb]
\renewcommand{\arraystretch}{0.9}
\footnotesize
\centering
\caption{Completion time (in milliseconds) of each process in \autoref{fig:flowchart}. Graphical results are also available in \cite{supp}.}
\label{table:time}
\begin{tabular}[b]{rrrrrrrr}
\cline{1-5} \cline{7-8}
$D$ & $n$ & a1 & a2 & a3 &  & b1 & b2 \\
\cline{1-5} \cline{7-8} 
10 & 100 & 0.011 & 0.006 & 0.003 &  & 0.914 & 0.002\\
10 & 1,000 & 0.014 & 0.010 & 0.016 &  & 4.417 & 0.002\\
10 & 10,000 & 0.067 & 0.091 & 0.144 &  & 42.357 & 0.002\\
\cline{1-5} \cline{7-8} 
100 & 100 & 0.029 & 0.004 & 0.022 &  & 0.900 & 0.110\\
100 & 1,000 & 0.072 & 0.010 & 0.160 &  & 4.410 & 0.110\\
100 & 10,000 & 0.949 & 0.085 & 1.578 &  & 42.338 & 0.110 \\
\cline{1-5} \cline{7-8} 
1,000 & 100 & 0.198 & 0.004 & 0.222 &  & 0.908 & 8.618\\
1,000 & 1,000 & 0.962 & 0.011 & 1.652 &  & 4.410 & 8.574\\
1,000 & 10,000 & 24.410 & 0.085 & 15.291 & & 42.335 & 8.580\\
\cline{1-5} \cline{7-8}
\end{tabular}
\end{table}

\begin{figure}[t]
	\centering
	\includegraphics[width=1.0\linewidth]{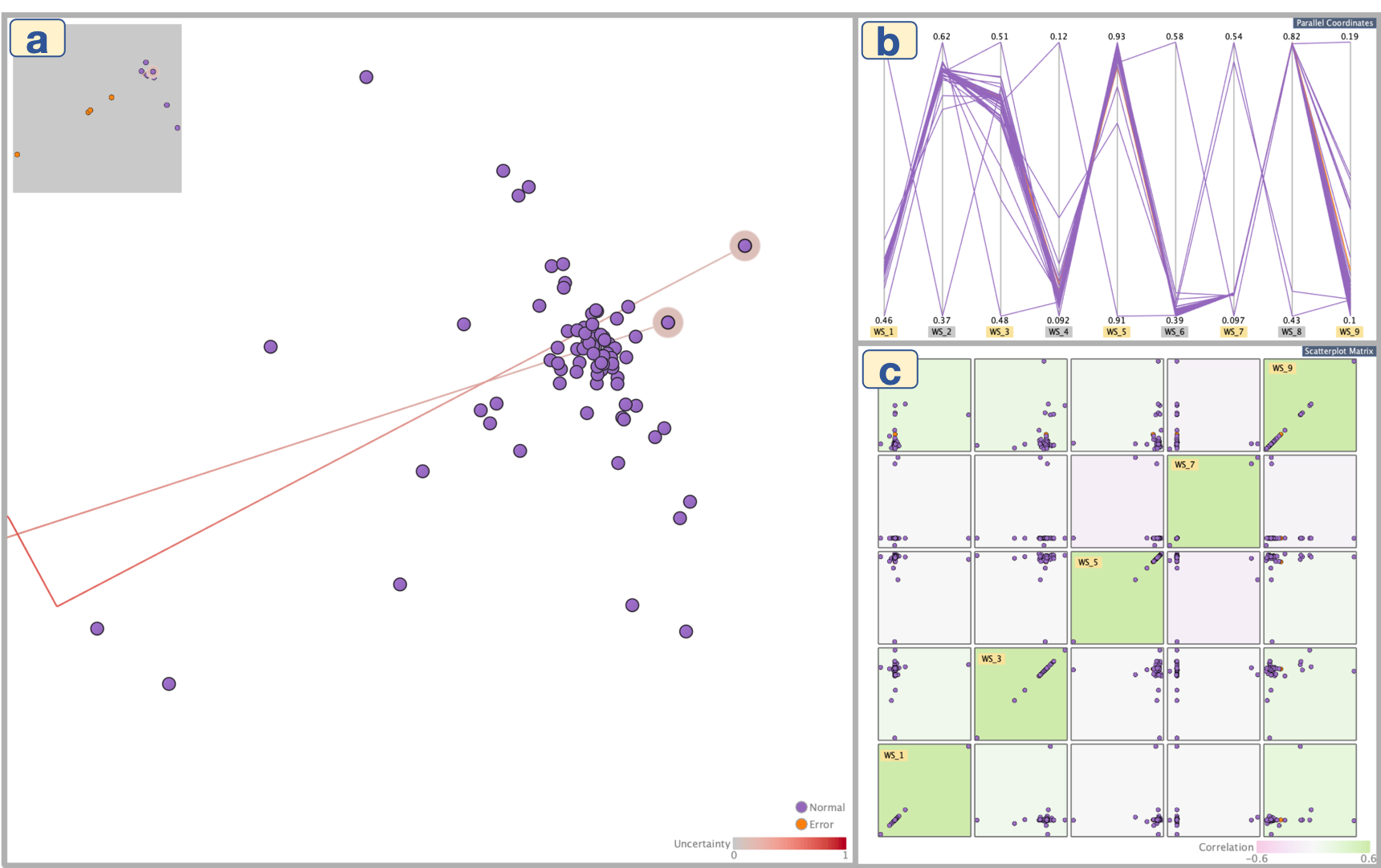}
  	\caption{A prototype system consisting of three views: (a) the DR view, (b) the parallel-coordinates view, and (c) the scatterplot-matrix view.}
	\label{fig:systemOverview}
\end{figure}

\begin{figure*}[tb]
    \captionsetup{farskip=0pt}
	\centering
    \includegraphics[width=0.9\linewidth,height=0.25\linewidth]{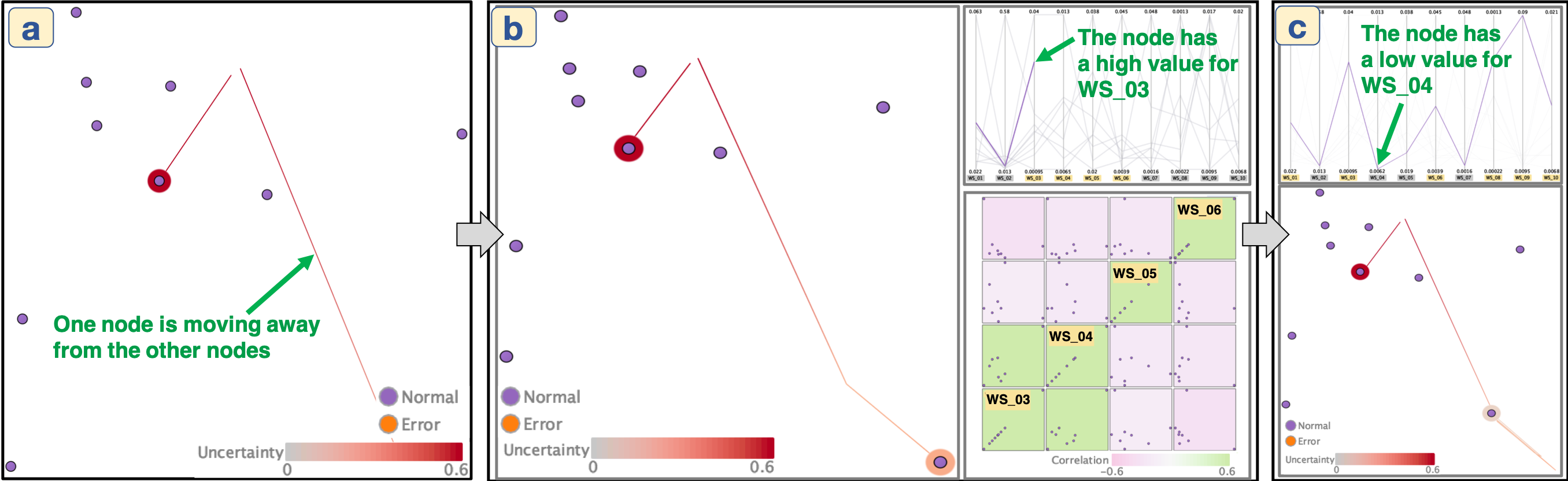}
   	\caption{An example of visual detection of an anomaly node. In (a), one node moves far away from other nodes and goes out of view from the current visualized range. This suspicious behavior indicates that this node could be an anomaly. Thus, we follow the node with our automatic detection in (b). To see more details about this node, we visualize it in the PCP view. We also show the scatterplots of the current work station (``WS\_03'') and three immediate work stations (``WS\_04'', ``WS\_05'', and ``WS\_06''). (c) shows the PCP and DR views after obtaining the entire set of values of the node.}
	\label{fig:caseStudy1_1}
\end{figure*}

\begin{figure}[tb]
    \captionsetup{farskip=0pt}
	\centering
    \includegraphics[width=0.83\linewidth,height=0.3\linewidth]{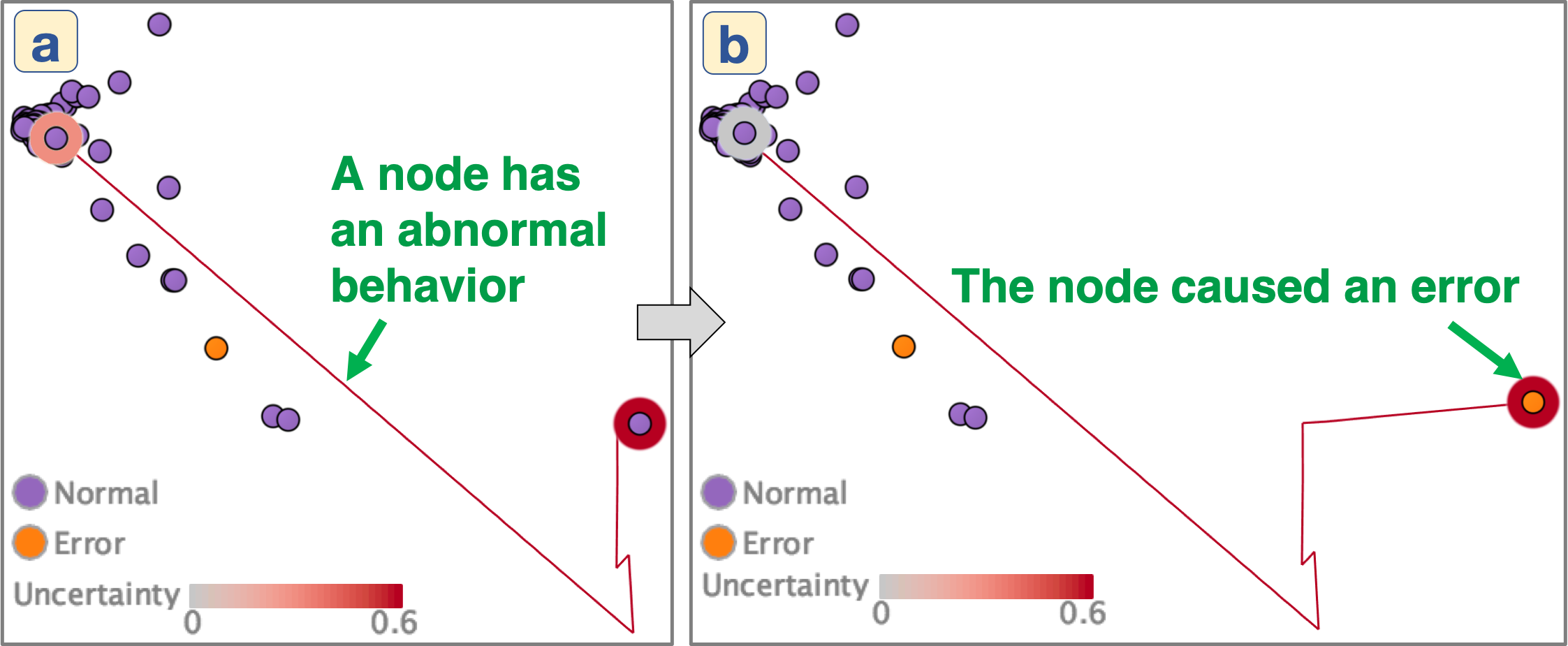}
   	\caption{An example of a visual prediction of a future error. In (a), we can see that one node has abnormal behaviors. At last, the node causes an error (indicated by orange color) as shown in (b).}
	\label{fig:caseStudy1_2}
\end{figure}

\autoref{table:time} shows the completion time for each method with different numbers of the pre-existing points ($n$=100,\,1,000,\,10,000). 
Each completion time is the average of ten executions.
In \autoref{table:time}, we can see that processes a1--a3 can be done in \SI{40}{ms} even when we have 10,000 pre-existing points with 1,000 dimensions, while b1--b2 can be done in approximately \SI{50}{ms}.
Note that the completion time for process a1 increases as $n$ increases even though incremental PCA's time complexity ($O(dm^2)$) does not relate to $n$. 
This is due to the projection step of ($n+m$) data points using the first $k$ principal components obtained from incremental PCA. 
These results show that the computational costs of our methods are low enough for supporting streaming data analysis with large numbers of data points and features in real-time.


\section{Prototype System}
\label{sec:prototype}
We develop a prototype system that integrates the methods described in \autoref{sec:methodology}. 
The prototype system has three views: the (a) DR, (b) parallel-coordinates (PCP), and (c) scatterplot-matrix (SM), as shown in \autoref{fig:systemOverview}. 
As the names indicate, the DR view shows the projection results from the incremental PCA, the PCP view displays the data points' values for each dimension with parallel coordinates, and the SM view presents the pairwise scatterplots between any of the two dimensions. 
While visualizing PCA results is effective in showing an overview of the streaming multivariate data, it neglects the detailed information of the data points. 
To supplement the DR view, we incorporate the parallel coordinates which can show many dimensions of information all at once in a limited space and reveal the trend of the data points clearly~\cite{cuzzocrea2013parallel}. 
However, the parallel coordinates are not suitable for analyzing the correlation between each pair of  dimensions~\cite{cuzzocrea2013parallel}. 
Thus, we provide a scatterplot matrix for this type of analysis. 
To achieve fast calculation and rendering, we use C++ and OpenGL for visualization, Qt for the user interface, and Eigen~\cite{eigen} for linear algebraic calculations.

In the DR view, we use the point's color to indicate a user-defined grouping of the data points.
For interactions, the system supports fundamental view operations and selection, such as zooming, panning, lasso selection, and filtering with linking to the other views.
From a dialog menu, the user can also start to use the automatic tracker described in \autoref{sec:autoTracking} with multiple options: track only incoming new data points, track only selected data points, or both. 

In the PCP view, each data point is shown as a polyline. 
The vertical axis corresponds to each dimension and its y-coordinates reflect the data points' values of its corresponding dimension.
The\,user\,can\,choose whether to scale the plotted values for each dimension\,within\,a\,0-to-1 range or not.
Each line color shows the corresponding group information defined in the DR view.
To perform brushing \& linking and filtering, we have provided a freeform selection for the parallel coordinates' lines.
The user can also select which dimensions to be shown in the SM view by clicking the names of the dimensions placed at the bottom of the view.
The selected dimensions are indicated in yellow. 

The SM view shows the pair-wise scatterplots between any two of the selected dimensions. 
To show the Pearson correlation coefficient for each plot, we use a colored background with a pink-to-green colormap (pink: negative correlation, green: positive correlation). 
This view also supports lasso selection. 


\begin{figure*}[tb]
    \captionsetup{farskip=0pt}
    \centering
    \includegraphics[width=0.85\linewidth,height=0.2\linewidth]{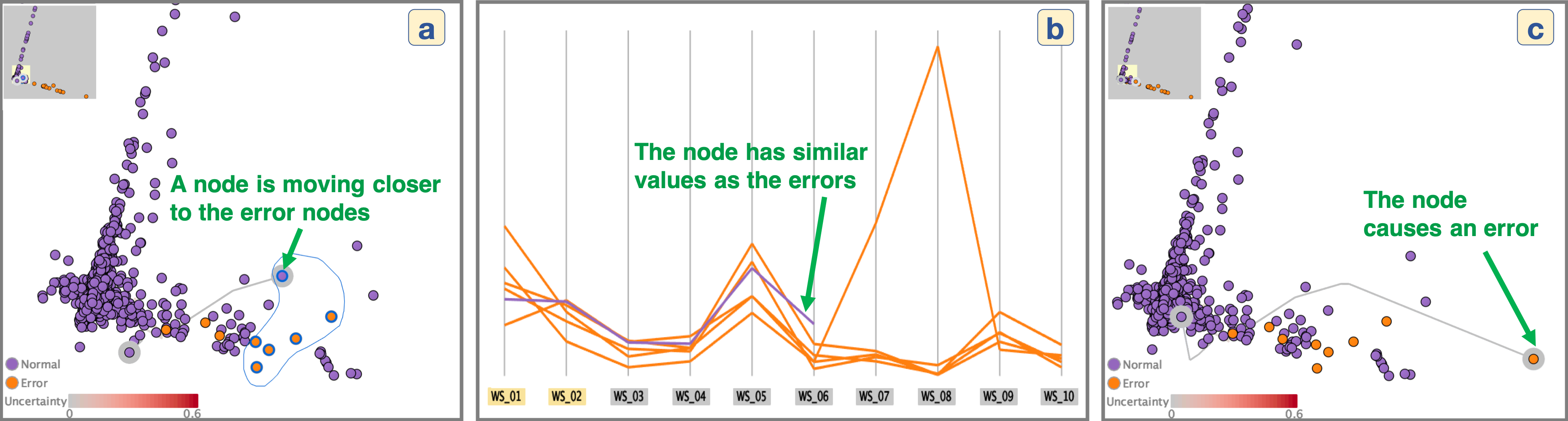}
    \caption{An example of a visual prediction of an error based on previous errors. In (a), we can see that one node, indicated with the green arrow, comes close to the error nodes colored with orange. From (a), we select this node and the error nodes with the lasso selection and visualized them as the parallel coordinates in (b). We can see that the values of the node (the purple polyline) follow closely to the values of the other nodes (the orange polylines). 
    At two work stations after the state of (a) and (b), the node causes an error, as shown in (c).
    }
    \label{fig:caseStudy1_3}
\end{figure*}

\begin{figure*}[tb]
    \captionsetup{farskip=0pt}
	\centering
    \includegraphics[width=1.0\linewidth,height=0.17\linewidth]{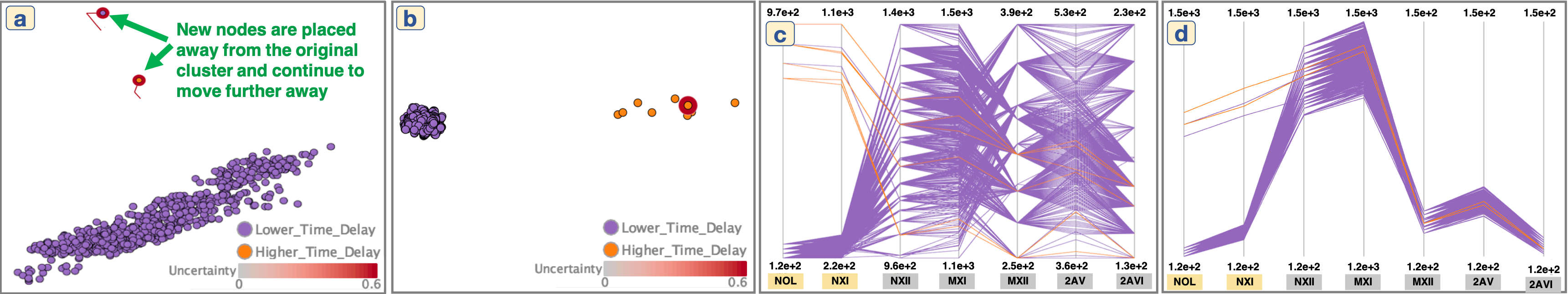}
   	\caption{An example of visual detection of a new forming cluster which has delays in the MTA bus trips. (a) shows two new nodes moved away from the originally formed cluster. (b) shows a distinct cluster (mainly consisted of orange nodes) formed with more arrived data points. (c) shows the PCP view of the two clusters. We can see that the values of the node (the purple polylines) show a clear distinction to the values of the new nodes (the orange polylines). 
   	   In (d), we show the PCP view with the original scales of the values. 
   	   }
   	\label{fig:caseStudy2_1}
\end{figure*}

\section{Case Studies}
We demonstrate the effectiveness of our incremental DR method for streaming multidimensional data with our prototype system.
By analyzing two different types of time-series data, we show how our method is used for finding useful patterns, such as anomalies and clusters.

\subsection{Visual Diagnosis of Assembly Line Performance}
We use real-time tracking data of an assembly line in a smart factory~\cite{xu2017vidx}. 
The assembly line consists of a set of work stations. 
Each product part is moving from one station to the next.
We use the status information sent from programmable logic controllers (PLCs) on the assembly lines when the parts arrive at the stations. 
We set each part at a work station as a data point and its cycle time as the data point's features.
The cycle time is calculated by subtracting the time range that a part finished the process of one station and has moved onto the next. 
There are 11 work stations in our selected subset of the assembly line. 
Therefore, if a part has finished passing through all 11 work stations, it has 10 features (cycle times).
In addition, we use the fault code to categorize the group information, which is recorded by the PLCs when any error occurs during processing a part on a station.

We now show three examples of visual diagnosis of anomalies and errors from a subset of the assembly line data in a single day. 
The full dataset consists of 1,728 product parts (data points), 10 cycle times (features) from 11 work stations, and the fault code (i.e., error or no error). 
The median of all cycle times from all parts is approximately one minute.

The first example is shown in \autoref{fig:caseStudy1_1}a, where we notice that one node (a product part) starts to move away from other nodes by looking at the path as indicated with a green arrow. 
Since this could be an anomaly, we start to track this node with our automatic tracking. 
As this node passes through more work stations, we find that this node keeps moving away from the other nodes, as shown in the DR view in \autoref{fig:caseStudy1_1}b.
To review in more detail, we select this node and show its data values in each dimension in the PCP view in \autoref{fig:caseStudy1_1}b.
We can see that this node has a high value for the work station ``WS\_03''. 
Then, we look at the scatterplot matrices for the work stations from ``WS\_03'' to ``WS\_06'' (the lower-right of \autoref{fig:caseStudy1_1}b). 
We can see that ``WS\_03'' has positive correlations with ``WS\_04'', while it has negative correlations with ``WS\_05'' and ``WS\_06''.
Therefore, if this node were to follow the same trend as the other nodes, we should expect that this node would have a high value for ``WS\_04'' and low values for ``WS\_05'' and ``WS\_06'', respectively. 
However, as shown with the PCP view at the top of \autoref{fig:caseStudy1_1}c, the node holds a low value for ``WS\_04''. 
Since the node behaves very differently from the other nodes, this foreshadows that this node will be an anomaly. 
As a result, as shown in the DR view at the bottom of \autoref{fig:caseStudy1_1}c, the node stays far away from the others. 
Despite the abnormal behavior, this does not cause an error during the process. 

As a second example, we show a case where a visually detected anomaly node causes an error. 
In \autoref{fig:caseStudy1_2}a, we find that one node suddenly strays away from the others and continues to move farther away in successive steps. 
Similar to the previous example, this behavior indicates a high possibility of the node being an anomaly. 
As a matter of fact, immediately after this step, the node causes an error, as shown in \autoref{fig:caseStudy1_2}b.
This example demonstrates the functionality of our method: to visually identify a data point which could cause an unknown error. 

In the third example, we demonstrate how we use our method to foresee a future error by utilizing the known errors.
As shown in \autoref{fig:caseStudy1_3}a, one node, as pointed by the green arrow, moves to a position where several error nodes (colored orange) reside. 
Since this behavior indicates that this node has a high possibility to cause the same error, we further investigate its relationships with those error nodes. 
We select the related nodes with a lasso selection in \autoref{fig:caseStudy1_3}a and visualize their values for each dimension with the PCP view, as shown in \autoref{fig:caseStudy1_3}b. 
From \autoref{fig:caseStudy1_3}b, we can see that the values of the node (represented with the purple polyline) have similar values with the error nodes (the orange polylines) up to the sixth work station.
Given this observation, we predict that this node will cause an error in the near future. 
In fact, we find that this specific node causes an error after it has passed two more work stations, as shown in \autoref{fig:caseStudy1_3}c. 

Through this case study, we find product parts that produce anomaly patterns and/or yield errors in the assembly line. 
We achieved this by applying our position estimation method on the product parts which have not passed all the work stations yet.
This shows the usefulness of our method to perform real-time monitoring on time-series data for early anomaly detection and error reasoning.


\subsection{Bus Traffic Analysis}
For the second case study, we use the tracking data from the Metropolitan Transportation Authority (MTA) and RTA (Regional Transportation Authority) at Nashville, United States~\cite{transitfeeds}. 
Nashville MTA/RTA Stops and Routes are used in mapping programs, such as Google Transit.
We use the arrival times of an MTA bus from one station to the next to calculate the transit time. 
Each data point in this dataset is a trip taken by each bus. The MTA dataset consists of many routes. For this case study, we pick one route that runs through downtown Nashville.
The dataset consists of approximately 1,500 data points (trips). 

\autoref{fig:caseStudy2_1} shows the results of processing 800 points. We can see that our incremental PCA has split the data points into one large, distinct cluster (purple points) as seen in \autoref{fig:caseStudy2_1}a. The newly incoming data points (the purple and orange points) promptly deviate from the large cluster. 
As more incoming points are processed, we see a new cluster forming, which mainly consists of orange nodes, as shown in \autoref{fig:caseStudy2_1}b. 
To understand how the clusters are being split, we further analyze the data with the PCP view, as shown in \autoref{fig:caseStudy2_1}c.
We find that the incoming orange nodes in the new cluster follow a different value on each bus stop when compared with the purple nodes. 
More specifically, we can see that these nodes have higher time delays for the first two stops (``NOL'' and ``NXI'') when compared to the purple nodes.
In \autoref{fig:caseStudy2_1}c, the parallel coordinates expand the values of each dimension to the minimum-maximum range on the $y$-direction. 
This makes judging more difficult on which cluster has higher delays in total. 
Alternatively, we choose to show the original values in each dimension without expanding them, as shown in \autoref{fig:caseStudy2_1}d.
This allows us to see that the new cluster does have a higher delay time in total. 
This example shows how we can discover and review newly emerging clusters with our method. 

\begin{figure}[tb]
    \captionsetup{farskip=0pt}
	\centering
    \includegraphics[width=0.85\linewidth,height=0.65\linewidth]{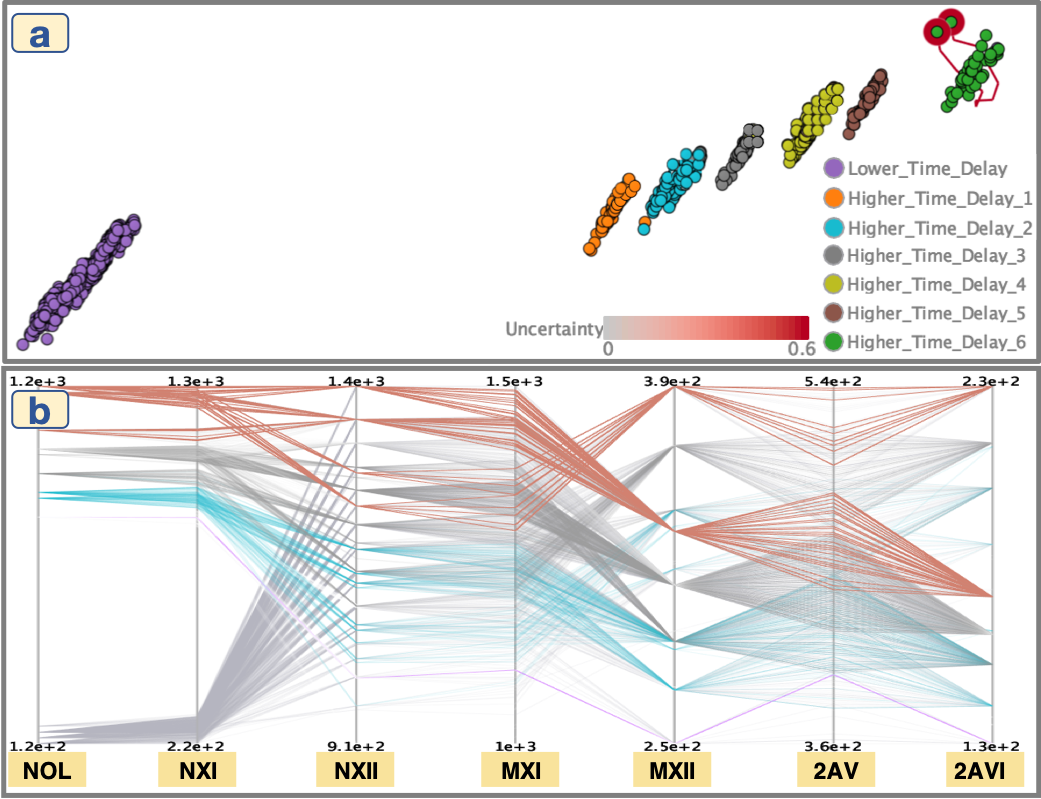}
    \caption{The bus trips after processing more data points from \autoref{fig:caseStudy2_1}.
    (a) shows the additional clusters formed away from the original cluster at \autoref{fig:caseStudy2_1}a. (b) shows the PCP view of the two selected clusters ``Higher\_Time\_Delay\_2'' (cyan) and ``Higher\_Time\_Delay\_5'' (brown). 
    }
    \label{fig:caseStudy2_2}
\end{figure}

As we continue to process more data points, referring to \autoref{fig:caseStudy2_2}, we observe that the data points with higher delays start to form more clusters. \autoref{fig:caseStudy2_2}a shows that there are six additional clusters being formed. Intuitively, we could assume that there is more correlation between the clusters that are closer to each other (i.e., the time delays are comparable). 
To understand which criteria leads to these separated clusters, we compare the values in each cluster with the PCP view. 
For example, in \autoref{fig:caseStudy2_2}b, we highlight the values of the clusters which mainly contain brown or cyan nodes by selecting from the DR view with the lasso selection. We can see that the brown lines (``Higher\_Time\_Delay\_5'') are mostly at the upper end of the figure when compared to the blue lines (``Higher\_Time\_Delay\_2''). 
With further investigation for each cluster, we find the clusters that are farther apart show higher time delays. 
Note, for the examples above, the coloring of the nodes is used to make the explanations clear. 
The findings and patterns can be found even by using the provided selection and filtering methods instead of using these colors.

\begin{figure}[tb]
    \captionsetup{farskip=0pt}
	\centering
    \includegraphics[width=0.85\linewidth,height=0.21\linewidth]{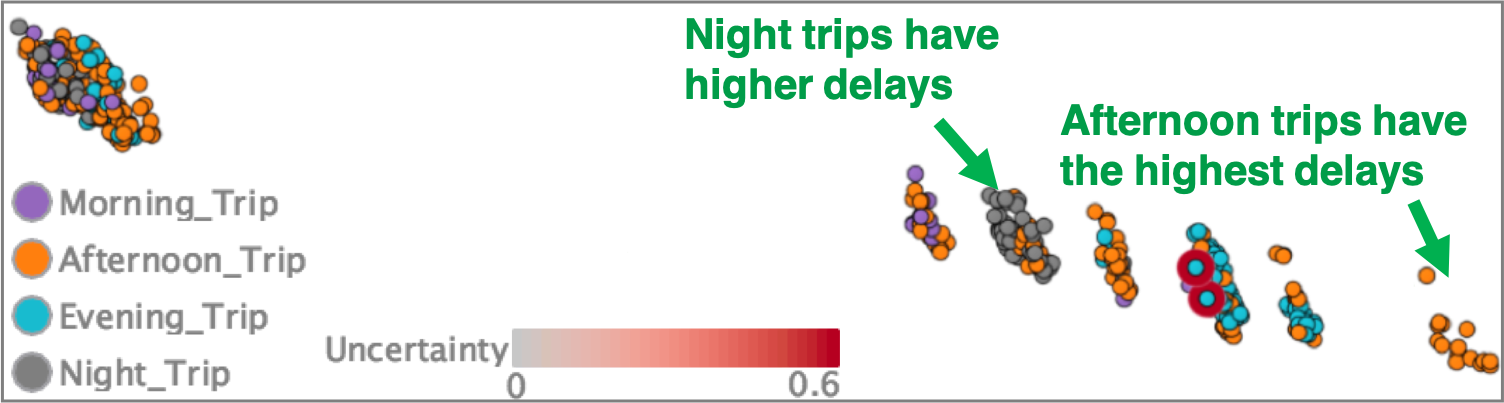}
   	\caption{The visually grouped bus trips colored by the hour of the day that each trip occurred in.}
	\label{fig:caseStudy2_3}
\end{figure}

At this point, we know how the clusters are formed. This information could be used to gain some additional insight from the dataset based on the time of the day when the bus trip occurred. 
We now then visually group the data points based on the hour of the day, and the groups are categorized as ``Morning''(5AM-11AM), ``Afternoon''(11AM-4PM), ``Evening''(4PM-9PM), and ``Night''(9PM-2AM) trips. 
From \autoref{fig:caseStudy2_3}, we notice that the highest delays (orange points in the middle right) occur in the ``Afternoon'' trips. 
In addition, the ``Afternoon'' trips' nodes can be found in all the clusters that have delays.
On the other hand, from our general assumption, one might assume that the ``Night'' trips would incur no delays.
However, this is not true, as seen in \autoref{fig:caseStudy2_3}.
Many of the ``Night'' trips are grouped into a cluster with delays (indicated with the green arrow in the middle). 
The time duration for the groups could be varied and this will change the final result of \autoref{fig:caseStudy2_3}. 
However, we consider that this particular choice gives us a concrete idea of how the data in our dataset is laid out in the final result.

\section{Discussion and Limitations}

To preserve the viewer's mental map, we use the Procrustes transformation consisting of translation, uniform scaling, rotating, and reflection. For the purpose of visualizing the data, the geometric transformations are not harmful because these transformations do not change the relative distance relationships among the data points. 
However, if the user wants to analyze the data based on the original PCA result, our algorithm can also provide enough information to restore the transformed result back to the original result.
This can be achieved by using $c$, $\boldsymbol{\tau}$ and $\mathbf{R}$ obtained with \autoref{eq:transform}. 
We would also like to note that the Procrustes transformation can be used to reduce the total positional changes between any two sets of data points (e.g., MDS results~\cite{gansner2009gmap} and node-link diagrams). 
This can help in the comparison of two different visualized results.

To deal with a non-uniform number of dimensions, we present the position estimation method utilizing the distance relationships among the new and exiting data points in the PCA result.
This approach is simple and generalizable, and thus can be applied to other incremental DR methods, such as an incremental MDS~\cite{williams2004steerable}. 
As described in \autoref{sec:varDim}, another potential option to handle a non-uniform number of dimensions is to predict the missing or unknown values using some machine learning approaches. 
Even though choosing a proper model for the prediction is challenging, it is worth pursuing as a following research.
With its predictive capability, our method can then possibly be used for streaming data with missing values in arbitrary dimensions.

\subsection{Limitations}

We employ the model by Ross et al.~\cite{ross2008incremental}, which requires at least two new data points to update the PCA result, as described in \autoref{sec:incPCA}.
When streaming data visualization requires frequent updates, this limitation is not a problem since, in most real-world scenarios, more than two new data points are constantly received. 
In cases where updates are not frequently occurred, we have enough time for updating the PCA result, and thus we can use the ordinary PCA instead. 
Also, our method inherits the limitation of a linear DR method and would not be suitable for revealing the local neighbors in the complicated structure.

In addition to the incremental addition of data points, our method allows the user to delete past observations by utilizing the forgetting factor in Ross et al.'s model~\cite{ross2008incremental}. 
However, our method does not support updating feature values of past observations because their model is not designed for such a case.
Exploring different ways to support this operation could be one direction for future work.

Another limitation of our work is related to the animated transitions. 
If we keep receiving new data points in a very short amount of time (e.g., less than a second), the staged animated transitions~\cite{bach2014graphdiaries} may not have enough time to complete. 
In this situation, we could consider not employing the animation. 
Despite that possibility, our method can still be effective in maintaining the mental map as it can keep the node positions with the geometric transformation. 
As an additional option, we can store the new data points for a period of time, and then update the result when there is enough duration for the animation. 

We design our position estimation method mainly for cases where new data points have an incomplete number of dimensions and keep collecting the values until they reach the same number of dimensions as the existing data points.
It is also possible to apply our method in other situations.
For example, when new data points have more dimensions than the existing data points, we can plot the new points by applying the incremental PCA with only the dimensions that the existing data points have. 
Even though some dimensions may be discarded, our uncertainty measure can be used to inform the user how much uncertainty is introduced. 
Another example is when some dimensions of the data are no longer able to be used (e.g., at some point a work station is removed from the assembly line).
By allowing the user to select which dimensions should be included in the PCA calculation, our method can also be applied for this case.

The scalability of the visualization is also worth discussing. 
In our prototype system, we visualize the PCA result as a 2D scatterplot. 
Therefore, the scalability issue mainly depends on the scatterplot itself.
One way we can approach this issue is to delete or aggregate data points which are not necessary to be visualized.
For example, we can filter out old data points in the visualization (e.g., data points that are a day old) or aggregate data points based on their similarities. 
This approach also solves the same scalability issue in the data points with the outer-ring colors.
A similar issue can occur when paths are drawn to show the new data points' movements.
If there are many new data points, it could create cluttered lines.
One way to reduce this clutter is to filter the paths.
For example, we can set a criterion based on the length of their movements or on a chosen threshold for specific feature values.

\section{Conclusions and Future Work}

DR methods are essential to many demanding data analysis tasks found in real-world applications. 
This work enhances the usability of a representative DR method for interactive analysis of streaming data.  
Our method is able to address both the interactivity and interpretability of the visualization. 
The visual stability and the capability of handling varying data dimensions offered by our incremental method lead to effective visualizations for streaming data analysis.
For example, our case studies demonstrate how time-varying data features, such as anomalies, errors, or clusters, could easily be more identified. 

We plan to extend our work within a few directions. 
First, we will investigate how to employ predictive methods for filling in the missing feature values. 
Incremental machine learning methods, such as online linear regression~\cite{strehl2008online}, can be used for achieving real-time model update. 
In addition, it is possible to customize the visualization based on specific analysis goals by interactively adjusting PCA~\cite{jeong2009ipca,sacha2017visual}; for example, we can support weighting of each dimension for PCA~\cite{jeong2009ipca}. Finally, we will extend to other DR methods (e.g., t-SNE) to support a variety of streaming data analysis tasks.

\acknowledgments{This research is sponsored in part by Bosch Research and the U.S.~National Science Foundation through grants IIS-1528203 and IIS-1741536.}

\bibliographystyle{abbrv-doi}

\bibliography{00_main}
\end{document}